\documentclass{article}
\usepackage{arxiv}

\usepackage{amssymb}
\usepackage{amsthm}
\usepackage{xcolor, soul}
\soulregister\cite7
\soulregister\autoref7

\definecolor{blue}{RGB}{41,5,195}
\definecolor{gray}{rgb}{.4,.4,.4}
\definecolor{gray}{rgb}{.4,.4,.4}
\definecolor{pblue}{rgb}{0.13,0.13,1}
\definecolor{pgreen}{rgb}{0,0.5,0}
\definecolor{pred}{rgb}{0.9,0,0}
\definecolor{pgrey}{rgb}{0.46,0.45,0.48}
\definecolor{lightgray}{rgb}{0.95, 0.95, 0.96}
\definecolor{whitesmoke}{rgb}{0.96, 0.96, 0.96}
\definecolor{javared}{rgb}{0.6,0,0} 
\definecolor{javagreen}{rgb}{0.25,0.5,0.35} 
\definecolor{javapurple}{rgb}{0.5,0,0.35} 
\definecolor{javadocblue}{rgb}{0.25,0.35,0.75} 

\usepackage{geometry}
\usepackage{graphicx}
\usepackage{caption}
\usepackage{subcaption}
\usepackage{adjustbox}
\usepackage{tabularx}
\usepackage{makecell}
\usepackage{tabto}
\usepackage{hyperref}
\usepackage{multirow}
\usepackage[section]{placeins}
\usepackage{setspace} 
\usepackage{amsmath}
\usepackage{listings}
\usepackage{lineno,hyperref}
\usepackage{float}
\newcommand\numberthis{\addtocounter{equation}{1}\tag{\theequation}}

\newcounter{bla}

\title{A fluid simulation system based on the MPS method}

\author{
 Andr\'e Luiz Buarque Vieira e Silva \\
  Voxar Labs - Centro de Informática\\
  Universidade Federal de Pernambuco\\
  Recife/PE 50740-560, Brazil. \\
  \texttt{albvs@cin.ufpe.br} \\
   \And
 Caio Jos\'e dos Santos Brito \\
  Voxar Labs - Centro de Informática\\
  Universidade Federal de Pernambuco\\
  Recife/PE 50740-560, Brazil. \\
  \texttt{cjsb@cin.ufpe.br} \\
  \And
 Francisco Paulo Magalh\~aes Sim\~oes \\
  Voxar Labs - Centro de Informática\\
  Universidade Federal de Pernambuco\\
  Recife/PE 50740-560, Brazil. \\
  [3pt]
  Departamento de Inform\'atica \\
  Instituto Federal de Pernambuco, Campus Belo Jardim \\
  Belo Jardim/PE 55145-065, Brazil \\
\texttt{fpms@cin.ufpe.br} \\
   \And
   Veronica Teichrieb \\
   Voxar Labs - Centro de Informática\\
  Universidade Federal de Pernambuco\\
  Recife/PE 50740-560, Brazil. \\
   \texttt{vt@cin.ufpe.br} \\
}

\begin{document}

\maketitle

\begin{abstract}
Fluid flow simulation is a highly active area with applications in a wide range of engineering problems and interactive systems. Meshless methods like the Moving Particle Semi-implicit (MPS) are a great alternative to deal efficiently with large deformations and free-surface flow. However, mesh-based approaches can achieve higher numerical precision than particle-based techniques with a performance cost. This paper presents a numerically stable and parallelized system that benefits from advances in the literature and parallel computing to obtain an adaptable MPS method. The proposed technique can simulate liquids using different approaches, such as two ways to calculate the particles' pressure, turbulent flow, and multiphase interaction. The method is evaluated under traditional tests cases presenting comparable results to recent techniques. This work integrates the previously mentioned advances into a single solution, which can switch on improvements, such as better momentum conservation and less spurious pressure oscillations, through a graphical interface. The code is entirely open-source under the GPLv3 free software license. The GPU-accelerated code reached speedups ranging from 3 to 43 times, depending on the total number of particles. The simulation runs at one fps for a case with approximately $200,000$ particles. Code: \url{https://github.com/andreluizbvs/VoxarMPS}

\end{abstract}


{\bf PROGRAM SUMMARY}

\begin{small}
\noindent
{\em Program Title: Voxar MPS}                                          \\
{\em Licensing provisions: GNU General Public License, version 3}                                   \\
{\em Programming language: C++ and CUDA}                                   \\
{\em Computer: Tested on CPUs: Intel Core i7-6820HK and Intel Core i7-7820HK and GPUs: GTX 1080 (mobile) and GTX 1080 Ti}                                   \\
{\em Operating System: Windows 10}                          \\
{\em CUDA: Tested on version 10.1 with driver version 445.87}                          \\
{\em Has the code been vectorised or parallelised: Different threads of CPU or number of cores of GPU}                          \\
{\em RAM: Tens of MB to several GB, depending on the scenario}                          \\
{\em Nature of problem: The Voxar MPS code has been developed to study the flow of incompressible fluids that requires high computational cost.}\\
{\em Solution method: Voxar MPS is an implementation of the Moving Particle Semi-implicit, a Lagrangian meshless particle method for incompressible fluids.}\\

\end{small}

\section{Introduction}
Some of the most common problems in naval hydrodynamics involve the study of fluid flow. For this, it is necessary to deal with large deformations near the surface, such as those presented in a good portion of computational mechanics problems \cite{cleary2006novel}. Also, Computer Graphics (CG) and Virtual Reality (VR) applications are constantly turning more realistic and interactive/real-time fluid simulations have been a frequent research topic in that area \cite{muller2003particle, daenzer2007real,bridson2015fluid}. Conventional techniques, as the Finite Element Methods (FEM) and Finite Difference Methods (FDM) are relatively inefficient when dealing with large deformations \cite{belytschko1996meshless} \cite{johnson1999advanced}. As an alternative, there are the Lagrangian meshfree or the particle-based methods. They achieve flexibility in situations where the classic techniques are too complex \cite{frey2005anisotropic}.

Two of the meshless techniques are the Moving Particle Semi-implicit method (MPS) \cite{koshizuka1996moving} and the Smoothed Particle Hydrodynamics (SPH) \cite{gingold1977smoothed} \cite{lucy1977numerical}, the latter initially intended to astrophysics applications and then adapted to fluid simulation. The MPS authors idealized it to simulate the flows of incompressible fluids, which refers to a fluid whose material density is constant within a fluid parcel, a property found in liquids. Its main difference from the original SPH, which can be considered an advantage for the MPS regarding numerical precision, is that the calculations adopt a semi-implicit predictor-corrector model \cite{koshizuka1996moving}. However, the SPH is more prevalent in CG and VR applications due to the high computational load occasioned by the MPS calculations, including solving the Poisson Pressure Equation (PPE).

In \cite{gotoh2016current}, the authors highlight the current achievements and future perspectives for projection-based particle methods, which are the ones that require solving a PPE. In it, the authors present a set of papers regarding the applicability of the MPS in ocean engineering, including wave breaking \cite{gotoh1999lagrangian} \cite{khayyer2008}, wave overtopping \cite{gotoh2005lagrangian}, wave impact \cite{khayyer2009modified} \cite{lee2011step}, green water on ships \cite{shibata2007numerical}, sediment transport \cite{gotoh2006key}, waves generated through landslide \cite{fu2015investigation} and fluid-structure interactions \cite{shibata2012lagrangian} \cite{hwang2014development}. 

One of the main issues of meshfree methods, in general, are in the spurious pressure oscillation of the particles \cite{khayyer2009modified}. Another disadvantage of particle-based approaches that rely on numerical precision is the application runtime \cite{zhu2011implementation}. Works found in the literature provide various kinds of performance speedups using a General Purpose Graphics Processing Unit (GPGPU) but always focusing on the standard MPS \cite{hori2011gpu, zhu2011implementation, taniguchi2014explicit}.

The system developed in this work allows combining numerical improvements to the MPS with different types of fluid flows. These include multiphase and viscoplastic simulations and parallel computation at different levels: multithreaded CPU and GPU. A fine-tuning of these combinations are available through a graphical user interface. It also allows the simulation of dam break cases, flood simulations, oil spilling disasters, and others with decent precision in less time. These features help to achieve general requirements in which precision and or time are somewhat relevant factors.

This research aims to provide a numerically enhanced MPS method implementation compared to its standard one, as well as expand the simulation tuning possibilities for different types of fluids and situations. Additionally, CPU and GPU parallelization are also provided, so more significant cases can reach interactive simulation and real-time rates. Modify and add new models to the system is straightforward. The resulting solution can assist in engineering problems regarding natural and environmental disasters in coastal and floodable areas. It can also be used in CG and VR, and, for some cases, real-time applications.

An unprecedented parallelization of an improved MPS version coupled with fluid models, such as turbulent and multiphase flow, is developed to run in CPU, through OpenMP \cite{dagum1998openmp}, and in GPU, by using CUDA \cite{cudatoolkit}. The tunable parameters are:

\begin{enumerate}
\item Multiple computational methods such as different pressure calculation approaches, different values for the kinematic viscosity, turbulent flow, and multiphase interaction; 
\item Improvements such as better momentum conservation, more precise discretization of differential operators and less spurious pressure oscillations; 
\item Choose between parallelized implementations using OpenMP and CUDA;
\item Interactive rates for simulations containing $10^5$ particles.
\end{enumerate}

The code is open-source, available to the community under the GNU General Public License v3.0 (GPLv3) license. There is also a README file with a detailed step-by-step installation description and information on how to use the developed system.

In the second section, there is a contextualization on the area, presenting modifications to the method over the years and acceleration techniques. Then, there is a section explaining the used technique and presenting a set of variations, improvements, and applications. Afterwards, there is an explanation regarding implementation, acceleration structures and user interface. Next, the main results are showcased, presenting each test case utilized, and their purpose. In the test cases section, a discussion is performed through a comparison with the results found in the literature. Subsequently, there is a performance analysis of CPU and GPU versions, memory usage, the speedups provided by the GPU parallelization, and the simulation frame rate. Finally, there is a discussion of final remarks and future works.

\section{Related works}
\label{sec:related}

Through the years, disasters involving natural phenomena have triggered research in different areas on how to avoid them. Fluid simulation focusing on liquids is one of these areas. To be able to simulate liquids coherently, meshless methods usually model the fluid flow as Weakly Compressible (WC) and Fully Incompressible (FI), which guarantees less fluid density oscillations. The MPS has also been providing significant assistance in that field since it intrinsically simulates incompressible flows \cite{chen2013frontiers}. The textbook chapter of Almeida et al. \cite{almeida2016meshless} is an introduction on the subject since it provides a thorough explanation of the MPS as well as some variations and implementation details. 

Despite the high level of applicability of the MPS, as shown in the previous section, there are only a few solutions or software that use this method, such as MPS-RYUJIN {\cite{mpsryujin}}, Particleworks {\cite{particleworks}} and MPARS {\cite{mpars}}. {Refs. {\cite{mpsryujin}} {\cite{particleworks}} are not open-source and Ref. {\cite{mpars}} does not provide any kind of parallelization, as opposed to the present work. Although there are other frameworks for fluid simulation, some of them being widely used, such as the DualSPHysics {\cite{crespo2015dualsphysics}} and GPUSPH {\cite{gpusph}}, they mainly focus on the SPH and its variations. 

\subsection{Modifications to the method}

Similar to the other meshfree methods, the MPS technique suffers from instability problems. Some of these issues are related to numerical errors at the boundaries, i.e., free-surfaces or solid boundaries interactions. Some works document and discuss the reasons for those problems \cite{kondoincompressible} \cite{lee2011step}. As an attempt to overcome these issues, some authors proposed changes to the method throughout the years; one of the most critical issues with the MPS is the spurious pressure oscillation. This subsection presents some of the works that successfully addressed those issues.

A set of papers by Khayyer \& Gotoh presents valuable insights and improvements to this problem, most of them proposing corrected differential operator models (Laplacian and gradient). \cite{khayyer2009modified} proposes modifications to the MPS to diminish spurious pressure fluctuations. The authors introduce a new formulation of the source term of the PPE, which is referred to as a Higher order Source term (HS), thus creating the CMPS-HS after combining this modification with their previous work. 
The authors in \cite{khayyer2010higher} focus on the Laplacian model used in the MPS. They derive a Higher order Laplacian model (HL) for the discretization of the Laplacian operator to refine further and stabilize the pressure calculation. Both Laplacian of pressure and the one corresponding to the viscous forces benefit from this model.
The authors remarked that, although the modifications improve pressure calculations, the numerical results still presented some unphysical numerical oscillation during tests.

Khayyer \& Gotoh \cite{khayyer2011enhancement} present two new modifications to resolve the shortcomings present in the method from their previous work. The first improvement deals with unphysical numerical oscillation caused by the source term in the PPE. Due to that issue, additional terms were added to it, referred by Error Compensating parts in the Source term of the PPE (ECS). The other change deals with situations with tensile instability \cite{MONAGHAN2000290}. It consists of a corrective matrix inserted in the pressure gradient calculations to achieve a more accurate approximation of the differential operator in question. The modifications here presented are going to be detailed further in \autoref{sec:methodology}.

In {\cite{gotoh2018on}}, the authors show the latest advances related to particle methods applied to coastal and ocean engineering, where they also include fully explicit methods. In one of them, Tayebi \& Jin \cite{tayebi2015development} proposes the Moving Particle Explicit (MPE) that solves an equation of state in a fully explicit form to obtain the particles' pressure. Other recent applications of the MPS brought up by the referred review are related to oil spilling {\cite{duan2017a}} and the swash beach process {\cite{harada2017numerical}}. In the present work, a fully explicit version of the MPS is studied, implemented, and accelerated.

\subsection{Acceleration}
Since the MPS is fully meshless, the particles are not connected explicitly by any edge, so, it is possible to optimize some computational aspects of the simulation, such as by parallelization, by cluster technology or General Purpose GPU (GPGPU) techniques \cite{vieira-e-silva2018improved}.
Shakibaeinia et al. \cite{shakibaeinia2010weakly} \cite{shakibaeinia2012mps} presented an alternative form of the MPS, where they incorporate more straightforward and effective formulations and fluid behaviours/flows models into the method. One of the advantages of doing so is to benefit from the MPS' numerically precise formulations without the main efficiency bottleneck, which is the Poisson pressure equation's solution.

Shakibaeinia and Jin \cite{shakibaeinia2010weakly} developed a technique based on the MPS interaction model in which a WC model replaces the FI one. By doing this, they decrease the necessary computation time. 
In Ref. \cite{shakibaeinia2012mps}, the authors extended their previous work by proposing a straightforward model of an immiscible multiphase method. They also employ and investigate different techniques for the viscosity model. In order to address turbulence issues in wave dynamics, they use the Large Eddy Simulation (LES) concept to formulate a sub-particle scale (SPS) turbulence model. The modifications mentioned above are studied and taken into consideration in the present work.

Hori et al. \cite{hori2011gpu} developed a GPU-accelerated version of MPS using CUDA. The authors focused on the search of neighbouring particles and the iterative solution of the linear system generated by the PPE, which generates a considerable computational load. They optimize the neighbouring particles search through a cell grid, in which there is a specific cell for each particle according to the particle's position. To compare accuracy and performance between the CPU and GPU-based codes, they execute 2-dimensional calculations of an elliptical drop evolution and a dam break flow. Finally, the reported speedup achieved is about 3 to 7 times. These speedup rates serve as comparison reference here.

In Ref. \cite{taniguchi2014explicit}, the authors focus on reducing the MPS runtime by replacing how the method calculates the particles' pressure. Instead of using a PPE, they implement an equation of state, similar to the work of Shakibaeinia and Jin \cite{shakibaeinia2010weakly}. It also aims to accelerate it by exploring its parallelization potential through a multi-core CPU, single-node GPU, and a multi-node GPU cluster environment. The authors use a domain subdivision approach to enable a simulation with a higher number of particles. For a 3D dam break test with $~700,000$ particles, the OpenMP solution could reach 5.3 times speedup, while the single-node GPU could reach 14.5 times speedup compared to a single-threaded CPU execution. The multi-node GPU with nine processes performs approximately 5.5 times faster than the single-node GPU. The authors claim that the proposed algorithm allows extensive WC-MPS simulations in distributed memory systems with reduced communication overhead. These speedup values are also a comparison reference to this work.

Unlike the works here presented, this work not only focuses on applying parallelization to the MPS calculations but also applying it to the added improvements and modifications. This combination of features is unprecedented. Besides, both parallelized versions (OpenMP and CUDA) can achieve speedups regarded to the MPS variations developed. Exemplifying, one set up of features can praise for stability and higher accuracy, while another may seek for higher performance. 

\section{The Moving Particle Semi-implicit Method}
\label{sec:methodology}
The MPS uses discrete elements called particles in which each of them carries a set of physical quantities. Since the fluid flow governing equations are for continuous domains, the continuous differential operators, such as the derivative, gradient, and Laplacian, need to be discretized. The MPS proposes discretized models to these operators. 
In this section, the method is detailed by showing its governing equations, discretized differential operations, and a set of variations and improvements to the standard MPS.

\subsection{Standard method \& Governing equations}\label{standardmethod}

Koshizuka and Oka \cite{koshizuka1996moving} models the fluid as a set of interacting particles, in which their motion is determined through the interaction with neighbouring particles and the governing equations of fluid motion. \autoref{eq1} and \autoref{eq2} are the continuity equation and Navier-Stokes equation, respectively, which describe the motion of one viscous fluid flow .

\begin{equation}\label{eq1}
\frac{1}{\rho}\frac{D\rho{}}{Dt}+\nabla{}\cdot \textbf{u}=0
\end{equation}

\begin{equation}\label{eq2}
\frac{\partial \textbf{u}}{\partial t} = - \frac{1}{\rho}\nabla{p} + \nu\nabla^2{\textbf{u}} + \frac{1}{\rho}\nabla \cdot \vec{\tau} + F_{ext} 
\end{equation}
where $\textbf{u}$ is the fluid velocity vector, $t$ is the time, $\rho$ is the fluid density, $p$ is the pressure, $\nu$ is the laminar kinematic viscosity, $\boldsymbol{\tau}$ is the sub-particle scale (SPS) or turbulence contributed by unresolved small motions (detailed in \autoref{subsec:turb}) and $F_{ext}$ represent external forces like gravity.

In this method, the domain is discretized into particles, as mentioned above. They interact with its neighbours through a kernel function $w(r)$, $r$ being the distance between two particles. A larger kernel size implies in an interaction with more particles. The system developed in this work provides a set of kernel functions to choose between them. However, it is recommended that the kernel function proposed by \cite{koshizuka1998numerical} should be used when running a PPE (FI version), as in \autoref{eq4}.

\begin{equation}\label{eq4}
w\left(\vert{}\textbf{r}_j-\textbf{r}_i\vert{}\right)=\left\{\begin{array}{l}\frac{r_e}{\left\vert{}\textbf{r}_j-\textbf{r}_i\right\vert{}}-1\
\ ,\ \ 0\leq{}\ r<r_e \\
0,\ \ \ r_e\leq{}r\end{array}\right.
\end{equation}
where $r_{e}$ is the radius of the interaction area and $\textbf{r}_i$ and $\textbf{r}_j$ are the positions of particles $i$ and $j$, respectively. When calculating the pressure through an equation of state (WC version), it is recommended to use and the kernel function proposed by \cite{shakibaeinia2010weakly}, as in \autoref{eq:wckernel}. Those pressure calculation models are addressed further in this section.

\begin{equation}\label{eq:wckernel}
w\left(\vert{}\textbf{r}_j-\textbf{r}_i\vert{}\right)=\left\{\begin{array}{l}\Big(1-\frac{\left\vert{}\textbf{r}_j-\textbf{r}_i\right\vert{}}{r_e}\Big)^3\
\ ,\ \ 0\leq{}\ r<r_e \\
0,\ \ \ r_e\leq{}r\end{array}\right.
\end{equation}

\autoref{eq5} defines $n_i$, the particle number density, at the particle's position $\textbf{r}_i$, which is proportional to the neighbors number of $i$.

\begin{equation}\label{eq5}
n_i=\sum_{j\not=i}w\left(\left\vert{}\textbf{r}_j-\textbf{r}_i\right\vert{}\right)
\end{equation}

The continuity equation is satisfied if the particle number density remains constant, and this constant value is denoted by $n_{0}$.

To identify a free-surface particle, the particle number density of the $i$th particle just needs to satisfy the condition presented in \autoref{eq15} since on the free-surface the particle number density drops abruptly.

\begin{equation}\label{eq15}
n_i<\beta{}n_0
\end{equation}
where $\beta$ is constant between $0.8$ and $0.99$. The bigger $\beta$ is, the bigger will be the number of particles recognized as free-surface. Koshizuka and Oka \cite{koshizuka1996moving} recommend to set it to $0.97$, and that is the value adopted here.

Generally, in meshless methods, there are two main approaches to calculate the particles' pressure when simulating liquids, the weakly compressible (WC) approach and the fully incompressible (FI) one, where each one of them has its advantages and disadvantages. This work implements both approaches, providing a wide variety of features so the users can decide based on their needs. After obtaining the particles' pressure, it is possible to compute the pressure gradient, enabling the calculations of the velocity values to update the particles' positions through a first-order Euler integration.

\subsubsection{Equation of state}

The WC model prioritizes performance since it severely diminishes computational load in exchange for numerical precision. In the work of \cite{shakibaeinia2010weakly}, the traditional FI model is replaced by a WC one because assembling and solving the PPE in each step takes a considerable amount of computation time: about two-thirds of each time step for a simulation with an order of magnitude of $10^{3}$ particles. The mentioned work replaces the PPE by an explicit relation, specifically an equation of state described by \cite{batchelor1970k} and modified by \cite{monaghan1994simulating} that is shown below.

\begin{equation}\label{eq16}
p_i^{k+1}=\frac{\rho{}c_0^2}{\gamma{}}\left({\left(\frac{n_i^*}{n_0}\right)}^{\gamma{}}-1\right)
\end{equation}
where $p_i^{k+1}$ is pressure value of particle $i$ in timestep $k+1$ and the typical value used for $\gamma = 7$. $c_{0}$ is the speed of sound. This study, in fact, shows a decrease in process time per time step while the simulation characteristics remains similar to the fully incompressible approach. Authors refer to this modified MPS as WC-MPS.

This work adopts the approach proposed by \cite{shakibaeinia2010weakly} due to its similarity to the weakly compressible approach adopted in the popular WCSPH \cite{monaghan2005smoothed}.

\subsubsection{Poisson pressure equation}

As opposed to the weakly compressible approach, the fully incompressible model calculates the particles' pressure with higher accuracy even though this leads to a high computational load. In this case, it is necessary to solve the Poisson Pressure Equation, shown in \autoref{eq13}, which yields a linear system of equations of the type shown in \autoref{eqlinearsystem}.

\begin{equation}\label{eq13}
\nabla{}{^2}p_i^{k+1}=-\frac{\rho{}}{\Delta{}t^2}\frac{n_i^*\ -n_0}{n_0}
\end{equation}
\begin{equation}\label{eqlinearsystem}
Ax = b
\end{equation}
where $A$ is a sparse square matrix of size $N \times N$ which N is the total particle number in the simulation, the Right-Hand Size (RHS) vector $b$ of size $N$ stores the source terms and $x$, also of size $N$, represents the desired pressures of the particles. 

The assemble of the coefficient matrix $A$ makes use of the discretized Laplacian model, shown in \autoref{eq7}.

\begin{equation}\label{eq7}
{\nabla{}}^2{\varphi{}}_i=\frac{2D_S}{n_0\lambda{}}\sum_{j\not=i}\left({\varphi{}}_j-{\varphi{}}_i\right)w\left(\left\vert{}\textbf{r}_j-\textbf{r}_i\right\vert{}\right)
\end{equation}

where $\varphi$ is some physical quantity, $\lambda$ is the weighted average of the squared distance between particles $i$ and $j$ (or $r_{ij}^{2}$), as shown in \cite{koshizuka1996moving}.

As previously mentioned in \autoref{standardmethod}, this work uses the ICCG to solve the PPE, like in \cite{koshizuka1996moving}. It consists of submitting the coefficient matrix $A$ into the incomplete Cholesky factorization, which is generally used as preconditioning for iterative numerical methods.
Afterwards, the conjugate gradient is applied to solve the linear system of equations iteratively. 

\subsection{Fluid flow models} \label{sec:behaviours}

The improvement of numerical stability and the addition of different models of fluid flow enable a more reliable simulation in certain situations. 
This section presents the used set of models of fluid flows in this work. 

\subsubsection{Multiphase flow}

A model that significantly increases the number of possible applications for the MPS is the one that supports multi-density fluids interaction, the multiphase flow.

Shakibaeinia and Jin \cite{shakibaeinia2012mps} proposed a straightforward multiphase model based on the MPS, in which it treats the system as a multi-density multi-viscosity fluid. The model is only applied to a WC-MPS \cite{shakibaeinia2010weakly}, solving a single set of equations for all phases. In this model, the density differences of particles of different phases are automatically taken care of, since that, when calculating a particle's velocity, its density appears directly in the equations. 

The main issue of this approach arises when dealing with the density discontinuity near the interface between the fluids, which can result in pressure field discontinuity. The strategy followed was to use a smoothed value of density $\langle \rho \rangle _i$ instead of the real particles' density, set for each particle before starting the simulation. 
The density of an individual particle is necessary for calculating its pressure. \autoref{eqmultiden} shows the multi-density pressure term in the weakly compressible model applied in this study.

\begin{equation}\label{eqmultiden}
\frac{1}{\rho_i} \left \langle \nabla p \right \rangle_i = \frac{d}{n_0} \sum_{j\neq  i}\bigg( \frac{(\rho_j/\rho_i) \alpha_j - \alpha_i}{r_{ij}^2} \mathbf{r}_{ij} w(\left | r_j - r_i \right |) \bigg)
\end{equation}
\begin{equation}
\alpha_i = \frac{c^2_0}{\gamma}((\left \langle n^* \right \rangle_i/n_0)^\gamma-1) 
\end{equation}
where $d$ refers to the number of dimensions in the simulation and the typical value of $\gamma = 7$ is used, as in Tait's equation of state \cite{batchelor2000introduction}.

The multiphase model proposed by Shakibaeinia and Jin \cite{shakibaeinia2012mps} can only be used by calculating the particles' pressure through the equation of state, which is the WC approach. Despite that, this work adopts this model for that incompressibility model, given its relative simplicity and stability. 

\subsubsection{Turbulent flow} \label{subsec:turb}

To calculate the influence of the turbulence term $\boldsymbol{\tau}$, referred to the unresolved small motion term in \cite{shao2005turbulence}, the large eddy simulation (LES) mathematical model for turbulence \cite{smagorinsky1963general} \cite{rogallo1984numerical} was employed.
According to the original LES conception, eddies capable of being resolved by the computational grid are allowed to evolve according to the Navier-Stokes equations, and a model is employed to represent the turbulence at sub-grid scales (mesh-based techniques). A sub-particle scale (SPS) model was made necessary for meshless methods. By introducing the turbulence eddy viscosity $\nu_t$, the unresolved SPS turbulence stress $\tau_{ij}$ in \autoref{eq2} can be written as shown in \autoref{eqturbstress}.

\begin{equation}\label{eqturbstress}
\frac{\tau_{ij}}{\rho} = 2\nu_t S_{ij} - \frac{2}{3}k\delta_{ij}
\end{equation}
where $\delta_{ij}$ is Kronecker's operator; and $S_{ij}$ is the strain rate and $k$ is the turbulence kinetic energy, which can be incorporated into the pressure term when solving the momentum equation \autoref{eq2}. The widely used model by \cite{smagorinsky1963general} is employed here to formulate the turbulence eddy viscosity.

\subsection{Numerical improvements} \label{subsec:enhancements}

In this section, there is a description of the implemented MPS variations.  It is noteworthy that the universe of MPS variations is vast, and the ones that were selected stand between improvement impact level and implementation cost. These variations allowed a version considered sufficiently stable and physically accurate to be achieved. It is also shown other modifications to the standard method, which expand the range of applications of the MPS.

\subsubsection {Momentum conservation}

A simple way to achieve consistent conservation of linear momentum is to ensure a better discretization of the gradient model. \autoref{eq18} shows the suggested alteration in the pressure gradient formulation by Khayyer and Gotoh \cite{khayyer2008development}.

\begin{align*}
\nabla{}\varphi_i=\frac{D_S}{n_0} \Bigg( \sum_{j\not=i}\frac{\left({\varphi_i+\varphi}_j)-({\hat{\varphi}}_i+{\hat{\varphi}}_j\right)}{{\left\vert{}\textbf{r}_j-\textbf{r}_i\right\vert{}}^2}
\left(\textbf{r}_j-\textbf{r}_i\right)w\left(\left\vert{}\textbf{r}_j-\textbf{r}_i\right\vert{}\right) \Bigg) \numberthis\label{eq18}
\end{align*}

When the anti-symmetric \autoref{eq18} is applied, linear momentum is conserved. This method is referred to by the authors as Corrected MPS (CMPS). 

\subsubsection {Pressure calculation}

One of the major issues of the MPS, and consequently widely explored, is the spurious pressure oscillation. Works that presented substantial improvements in this area, making few and simple modifications to the method, have been proposed \cite{khayyer2009modified, khayyer2010higher}. The authors call the first one the MPS with a Higher order Source term (MPS-HS) since it presents a new formulation for the calculation of the derivative of the particle number density $(\frac{Dn}{Dt})$. Using this variation, the \autoref{eq13} is replaced by the \autoref{eq21} \cite{khayyer20123d}.

\begin{equation}\label{eq21}
\nabla{}{^2}p_i^{k+1}=-\frac{\rho{}}{n_0\Delta{}t} \left(\sum_{i\not=j}\frac{r_e}{r_{ij}^3}\left(x_{ij}u_{ij} \right. \right. + \left. \left. y_{ij}v_{ij}+z_{ij}w_{ij})\Bigg)^* \right. \right.
\end{equation}
where $r_{ij}$ is the distance between particles $i$ and $j$. $x_{ij}$, $y_{ij}$ and $z_{ij}$ represent the distance between particles $i$ and $j$ in each dimension and $u_{ij}$, $v_{ij}$ and $w_{ij}$ the velocity difference of particles $i$ and $j$ in each dimension.

Another improvement to the implemented pressure calculation was the proposition of a higher order Laplacian model for both two and three (\autoref{eq23}) dimensional simulations \cite{khayyer2010higher, khayyer20123d}.

\begin{equation}\label{eq23}
{\nabla{}}^2{\varphi{}}_i=\frac{1}{n_0}\sum_{i\not=j}\left(\frac{2{\varphi{}}_{ij}r_e}{r_{ij}}\right)
\end{equation}
where $\varphi$ is a generic physical quantity. This new derivation was named by the authors as MPS with a Higher order Laplacian of pressure (MPS-HL).

\subsubsection{Numerical stability}

Khayyer and Gotoh \cite{khayyer2011enhancement} came up with a PPE's source term with error-compensating parts to enhance even further pressure and velocity field calculations. The compensating parts should be measures for instantaneous and accumulative violations of fluid incompressibility. \autoref{eq24} shows the new terms, and \autoref{eq25} shows the complete modified pressure calculation equation.

\begin{align*}
ECS=\left\vert{}\left(\frac{n^k-n_0}{n_0}\right)\right\vert{}\left[\frac{1}{n_0}{\left(\frac{Dn}{Dt}\right)}_i^k\right]
 + \left\vert{}\left(\frac{\Delta{}t}{n_0}{\left(\frac{Dn}{Dt}\right)}_i^k\right)\right\vert{}\left[\frac{1}{\Delta{}t}\frac{{n^k-n}_0}{n_0}\right]
\numberthis\label{eq24}
\end{align*}

\begin{equation}\label{eq25}
\nabla{}{^2}p_i^{k+1}=\frac{\rho{}}{n_0\Delta{}t}{{\left(\frac{Dn}{Dt}\right)}_i}^*+ECS
\end{equation}

According to Gotoh \cite{gotoh2013advanced}, this method ensures satisfactory accuracy and stable computation, more specifically, under the absence of tensile stress. Ref. \cite{gotoh2013advanced} shows a comparison between it and the standard MPS. 

\section{Implementation}
\label{sec:dde}

This work implements the MPS through the C/C++ programming languages. OpenMP \cite{dagum1998openmp} and CUDA \cite{cudatoolkit} were used to take advantage of the many cores in the CPU and GPU, hence accelerating the program execution. Lastly, there is a graphical user interface (GUI) to aid in the system usage, developed through the Windows Forms graphical class library from the Microsoft .NET framework \cite{platt2002introducing}.

\subsection{Neighbouring search algorithms}
\label{subsec:neigh}

This study adopts the "cell-linked list" strategy for the neighbourhood search \cite{shakibaeinia2010weakly}. It relies on a background Cartesian grid that divides the whole domain into cells. They have sides equal to $r_e$, which is the influence radius of a particle. In every iteration, the particles of the simulation are allocated in a specific cell, depending on their position. Thereby, when searching for a particle's neighbours, it is only necessary to look in the cell of the particle itself and adjacent cells. Thus, there is a list of particles that remain constant for that entire step, which offers a considerable decrease in the complexity of the neighbouring search function to $O(n \ log \ n)$.

\subsection{OpenMP}

The machine used in the present study holds an Intel\textsuperscript{\textregistered} Core\texttrademark~i7-6820HK CPU @ 2.70 GHz \cite{inteli74790} with 32 GB of installed RAM, a 64-bit operating system (x64), with 4 cores. 
\autoref{lst:openmp} shows an example of the particle number density function main loop parallelized just by adding one line of code, highlighted in red (line 1).

\begin{lstlisting}[float,caption= {C code of the particle number density calculation benefiting from OpenMP},frame=single, label={lst:openmp}, moredelim={[is][\color{red} \textbf]{@}{@}}]
@#pragma omp parallel for schedule (guided)@
for (int i = 1; i <= num_of_particles; i++) {
    double sum = 0.0;
    for (int l = 2; l <= neighb[i][1]; l++) {
        int j = neighb[i][l];
            (...)   /*sum of kernel values for particle i and each of its neighbors j*/
    }
    n[i] = sum;
}
return(n);
\end{lstlisting}

\subsection{CUDA}

The GPU code was developed based on the fully sequential and the OpenMP versions previously presented using CUDA C/C++. The utilized machine has an NVIDIA GeForce GTX 1080 Ti, which contains a total of $3,584$ CUDA cores and  11 GB of video memory. The number of threads per block is set to 256. As for the implementations, while the parallelization process of some functions was straightforward, some others needed adaptations for a parallelized version. 

A tricky parallelization step in the fully incompressible model of the MPS is the PPE solution. For this work, the standard ICCG solver used by Koshizuka and his colleagues \cite{koshizuka1998numerical} was parallelized in both OpenMP and CUDA versions using raw array formats for the linear system elements, such as the coefficient matrix and right side source vector. Also, to decrease memory usage by data structures, the total number of possible neighbours ($N_{neigh}$) of a particle was limited to $300$, which is the value used in \cite{koshizuka1996moving}. Doing so allowed assembling a coefficient matrix with $N \times N_{neigh}$ elements with $N$ being the total number of particles, rather than an $N \times N$ matrix. \autoref{lst:ppesol} shows the PPE assembly in a CUDA kernel.

\begin{lstlisting}[float, caption= {PPE assemble GPU code},frame=single, label={lst:ppesol}, moredelim={[is][\color{red} \textbf]{@}{@}}]
@unsigned int i = offset + (blockDim.x * blockIdx.x + threadIdx.x);@
double val = 0.0;
if (bcon[i] == 0) { // if fluid or boundary p. 
    poiss[i*num_of_neighbors + 1] = 0.0;
    for (int l = 2; l <= neigh[i*num_of_neighbors + 1]; l++) {
        int j = neigh[i*num_of_neighbors + l];
        if (bcon[j] == -1) poiss[i*num_of_neighbors + l] = 0.0; //if dummy p.
        else {
            /* Calculating distance `dist' between particles ... */
            (...)
            /*Calculating value `val' for the matrix assemble ...*/
            poiss[i*num_of_neighbors + l] = -val;
            poiss[i*num_of_neighbors + 1] += val;
        }
    }
    poiss[i*num_of_neighbors + 1] += (1.00e-7) / time_step / time_step;
}

\end{lstlisting}

Another significant step of the method, performance-wise, is the neighbourhood search function. This step occupies a large portion of the simulation running time in either pressure calculation models, making its parallelization essential to the practicability of massive simulations. In this work, the implementation of the parallelized neighbourhood search is also based on the cell linked list approach described in \autoref{subsec:neigh}, but, its parallelization process is based on the one from Ref. \cite{tavker2018parallel}, which presents a high scalability capacity.

\subsection{Graphical user interface}

The user interaction may happen through a GUI. \autoref{fig:gui} shows its layout and design. In it, (1) is a combo box from which the user can choose whether the simulation will have two or three dimensions; in (2) the user can choose how the code will run: sequentially, in parallel in CPU (through OpenMP) or parallel in GPU (through CUDA); in (3) the user will select a previously built simulation scenario; in (4) two approaches of pressure calculation can be chosen: weakly compressible or fully incompressible. By selecting the latter, (5) and (6) will be available, which are options that, if checked, enable Khayyer and Gotoh \autoref{subsec:enhancements} models of pressure gradient calculations to better conserve the linear and angular momentum of the fluid flow. The checking of (7) employs the SPS-LES turbulence model; (8) is only available if the test scenario chosen allows multiphase interaction. If checked, it enables viscoplastic properties in the second fluid in the simulation. (9) and (11) are the density values [$Kg/m^3$] of the fluids in the simulation, and (10) and (12) are their kinematic viscosity [$m^2/s$]; (13) sets the time duration between two steps of the simulation and (14) sets how long it will last, both in seconds. If checked, (15) creates a folder in the executable file directory containing all the particles' information in each time step. In contrast, (16) creates a folder with the fluid particles' information in each time step. (17) starts generating the simulation, and (18) allows the user to switch between languages.
 
\begin{figure}[ht]
\centering
\includegraphics[width=1.0\linewidth]{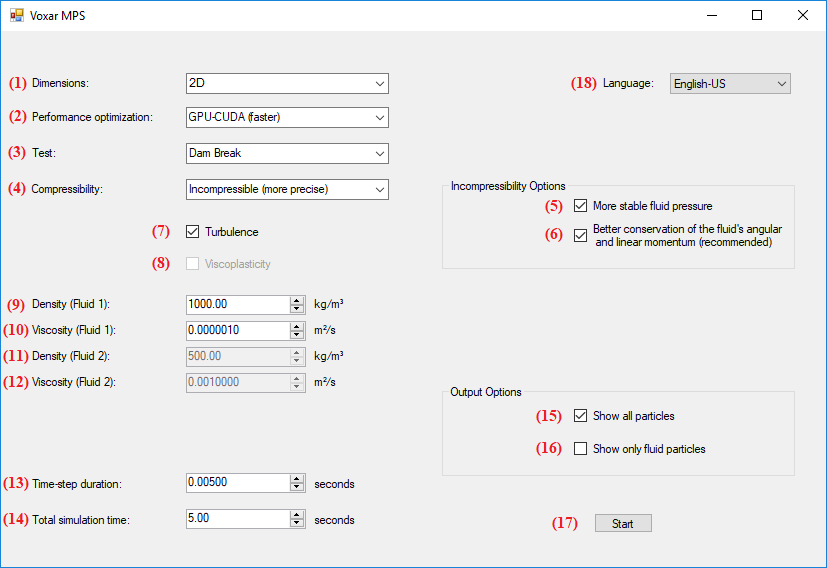}
\caption{\label{fig:gui}  User interface main screen}
\end{figure}

\section{Evaluation \& discussion}
\label{chap:evaluation}

The aim here is to validate the models and properties of the developed method, such as incompressibility, numerical precision, turbulent flow, and multiphase interaction. Another goal in this section is to assess the performance gain and frame rates of the parallelized implementations of the weakly compressible and fully incompressible versions of the MPS and discuss them.

\subsection{Water drop}

The examination of the evolution of an elliptic water drop is a common test case to validate incompressibility models of a particle-based fluid simulation method \cite{monaghan1994simulating} \cite{bonet1999variational} \cite{shakibaeinia2010weakly}. The usual test consists of a two-dimensional water drop, beginning the simulation in the shape of a circle, with a predefined velocity field of $(-100x, 100y)~m/s$ so that its format evolves into an elliptical shape over time. \autoref{fig:dropcase} shows a sketch of the test's geometry. The water drop has a circle radius of $9\times10^{-2}~m$ and an average particle distance of $5 \times 10^{-2}$ $m$, implying a total of $1257$ particles. The influence radius is $ r_e = 3l_0$, where $l_0$ is the average particle distance, and the adopted time  was $10^{-5}~s$. The used configuration for this test was the default one, as presented in \autoref{fig:gui}, with a fluid density equal to $10^3~Kg/m^3$ and fluid viscosity $10^{-6}~m^2/s$, only differing in the field (3) since it is the Water drop test.

\begin{figure}[ht]
\centering
\includegraphics[width=0.35\textwidth]{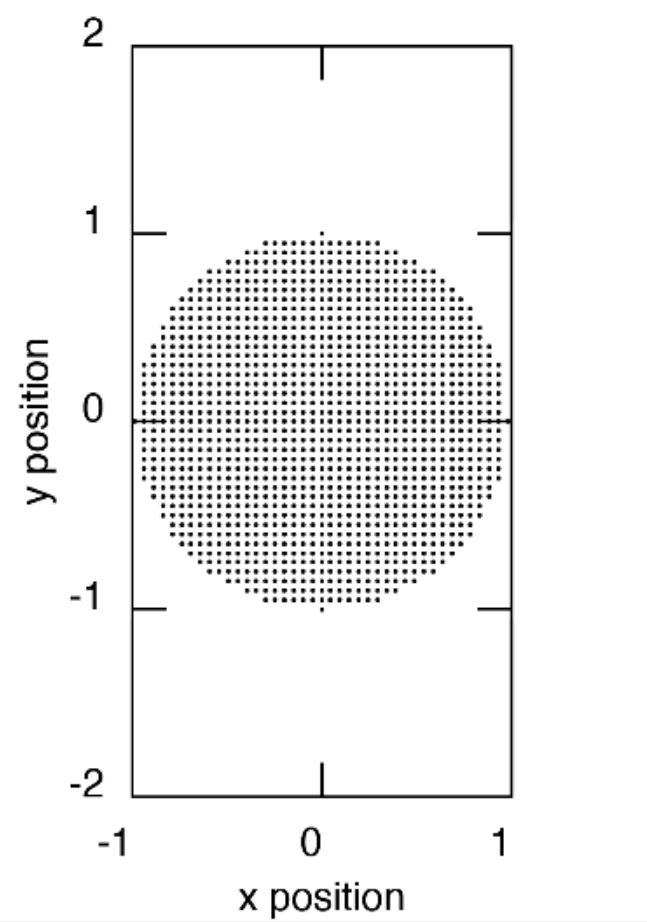}
\caption{\label{fig:dropcase} Initial state for the simulation in the water drop test (scale in meters)}
\end{figure}

As in \cite{shakibaeinia2010weakly}, \autoref{fig:dropwatersteps} shows three time instants of the simulation. The kernel function used here is in \autoref{eq:wckernel}.

\begin{figure}
\centering
\begin{subfigure}{.33\linewidth}
  \centering
  \includegraphics[width=.99\linewidth]{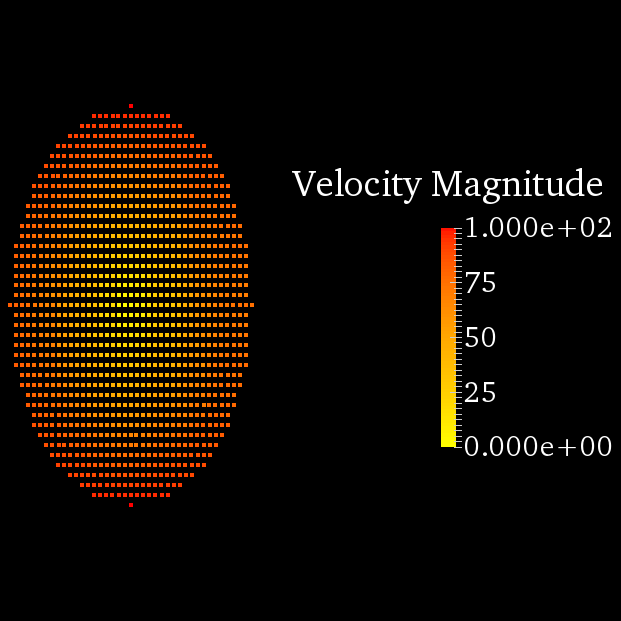}
  \caption{ $t=0.0025~s$ }
  \label{fig:drop0.0025}
\end{subfigure}%
\begin{subfigure}{.33\linewidth}
  \centering
  \includegraphics[width=.99\linewidth]{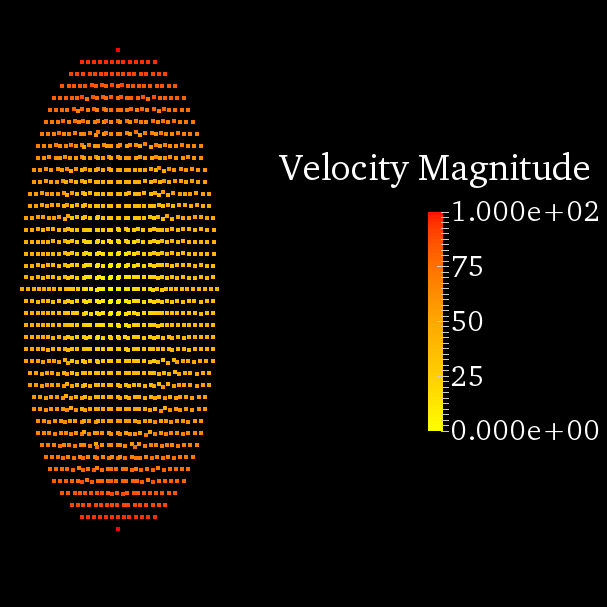}
  \caption{$t=0.0050~s$}
  \label{fig:drop0.0050}
\end{subfigure}
\begin{subfigure}{.33\linewidth}
  \centering
  \includegraphics[width=.99\linewidth]{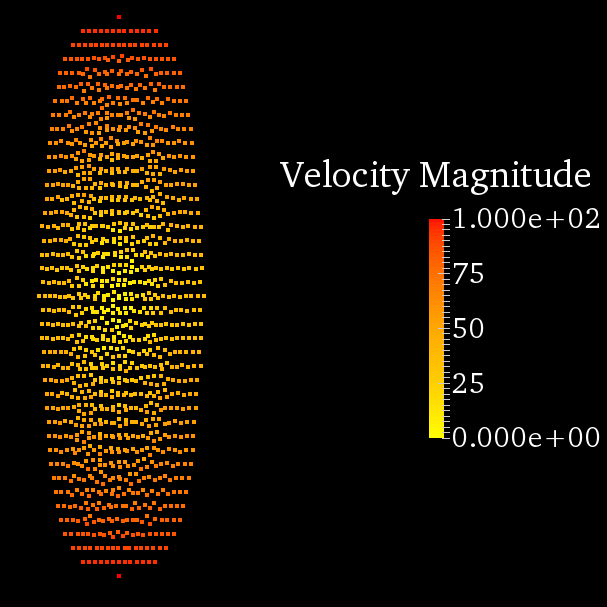}
  \caption{$t=0.0075~s$}
  \label{fig:drop0.0075}
\end{subfigure}

\caption{Velocity magnitude profiles (in $m/s$) of the water drop in three different time instants}
\label{fig:dropwatersteps}
\end{figure}

As noted by Monaghan \cite{monaghan1994simulating}, the condition in this test to measure incompressibility is that $ab$ is constant throughout the simulation, where $a$ is the semi-minor axis, and $b$ is the semi-major axis of the ellipse. This simulation ends when the size of $b$ becomes twice the value it was initially, and, in that instant, the $ab$ value is compared to its initial value. Errors are within less than $0.3~\%$. \autoref{dropdata} compares this result to \cite{monaghan1994simulating} which keep errors in less than $2~\%$ and \cite{shakibaeinia2010weakly} which keep errors in less than $0.2~\%$. This comparison shows the achieved result matches those from previous works.

\begin{table}[ht]
\centering
\caption{Comparison of water drop dimensions throughout the simulation}
\label{dropdata}
\begin{tabular}{cccccc}
        & \vline & $t=0.0~s$ & $t=0.01~s$ & \space & Difference in \% \\ \hline
$b$ (meters) & \vline & $0.875$  & $1.75$    & \space & $100$             \\
$ab$ (meters) & \vline & $0.766$  & $0.764$   & \space & $0.26$               
\end{tabular}
\end{table}

\subsection{Dam break}\label{sec:dambreak}

The collapse of a water column has been widely used in the literature to validate the numerical precision of various fluid simulation techniques. 

As the test performed by \cite{koshizuka1996moving}, in this work, the water column height is two times bigger than the water column length $L$. The floor in the model employed here is also four times the length of the water column. The size of the water column varies depending on how many particles the simulation has. The average particle distance is $10^{-2}~m$, and the time step of the simulation is $5 \times 10^{-3}~s$. The total number of particles is $1122$. 

The authors of the original MPS, referred here as standard MPS, put it to test by comparing it to a volume of fluid (VOF) approach \cite{hirt1981volume} and experimental data from three different experiments \cite{koshizuka1995particle,martin1952experimental}. This test adopts the standard dam break model \cite{koshizuka1996moving} and compares the simulation results obtained with the experimental data and the other simulation results. 
This comparison is possible by examining the water leading-edge position over time, since the dam burst until it hits the right wall. The leading edge is the front of the collapsing water column running on the floor (bottom wall).

\autoref{fig:koshiandmine} compares directly experimental data, other methods results, with the proposed technique. The configuration used here is the default as well, as depicted in \autoref{fig:gui}. The leading edge position is dimensionless, $Z/L$, where $L$ is the water column's initial length, which for this test is equal to $0.18~m$. If the water column size changes, the value of $0.18~m$ also changes, respecting the average particle distance $l_0$. The time axis in the chart is dimensionless as well, $t \sqrt{2g/L}$, where $t$ is the time in seconds and $g$ the gravitational acceleration, which is equal to $9.8~m/s^2$. 

\begin{figure}[ht]
\centering
\includegraphics[width=0.5\linewidth]{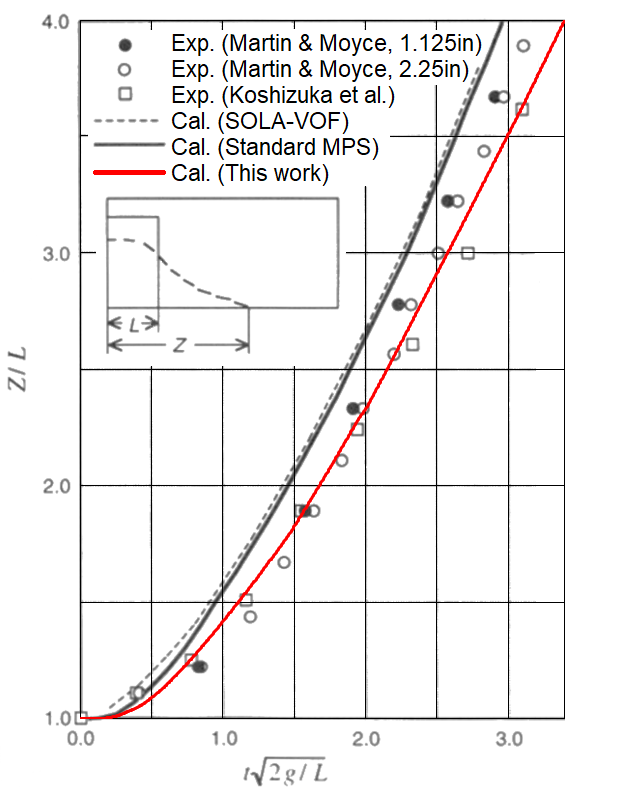}
\caption{\label{fig:koshiandmine} Direct comparison of the evolution of leading edge position between the developed technique (in red), and the experimental results and other methods. Background image extracted from \cite{koshizuka1996moving}}
\end{figure}

\autoref{fig:koshiandmine} shows that the developed method gets a lot closer to the experimental data than the others. Interestingly, it almost overlaps the experiment represented by the small empty squares from \cite{koshizuka1995particle}. It is important to note that, differently from a technique, the experiments do not become outdated since the setup did not change over the years. Therefore, this corroborates the assumption of a higher precision of the implemented method in this study.

\subsection{R-T instability}

The Rayleigh-Taylor instability test's goal is to show the multiphase model ability to handle the density stratification and to evolve a linear perturbation into nonlinear hydrodynamic turbulence \cite{youngs1984numerical}. 

The test consists of placing the same amount of two immiscible fluids with different densities and the same kinematic viscosity, one on the top of the other. The heavier one will stay on top of the lighter, only influenced by the gravitational force $g$ inside a two-dimensional rectangular box. When the simulation starts, the denser fluid will tend to go downwards, pushing the lighter upwards. The interface between the fluids will become unstable, and the format of the pattern generated at the interface will say whether the hydrodynamic turbulence was formed or not. The lighter fluid forms a bubble in the shape of a mushroom cap that breaks eventually. The Atwood number, calculated as shown in \autoref{eq:atwood}, characterizes the problem.

\begin{equation}\label{eq:atwood}
A_t=\frac{\rho_h - \rho_l}{\rho_h + \rho_l}
\end{equation}

$\rho_h$ and $\rho_l$ refer to the heavier and lighter fluid densities, respectively. For the test used in this study, $A_t = 1/3$ and the kinematic viscosity $\nu = 0.010$ for both fluids, following \cite{shakibaeinia2012mps}. The test has $3066$ particles. The system configuration used in this scenario differs from the default options shown in \autoref{fig:gui} with some differences. Here, field (3) is R-T instability, (4) is Weakly compressible, and fields (9) and (11) respect the Atwood number calculation. \autoref{fig:rt-insta-our} shows some steps of the generated simulation. Ref. \cite{shakibaeinia2012mps} shows a similar test displaying the same phenomenon. For a better context on how other particle-based approaches behave, their results are shown next to the generated simulation in \autoref{fig:rt-insta-shaki} . The main difference between them is that, in the scenario from Ref. \cite{shakibaeinia2012mps}, there is already an initial perturbation to the fluids' interface. Another difference is that in the test performed here, the domain top is open. These differences influence the test, causing the bubble (mushroom cap) to be upside-down.

\begin{figure}[ht]
\begin{subfigure}{\linewidth}
  \centering
  \includegraphics[width=.8\linewidth]{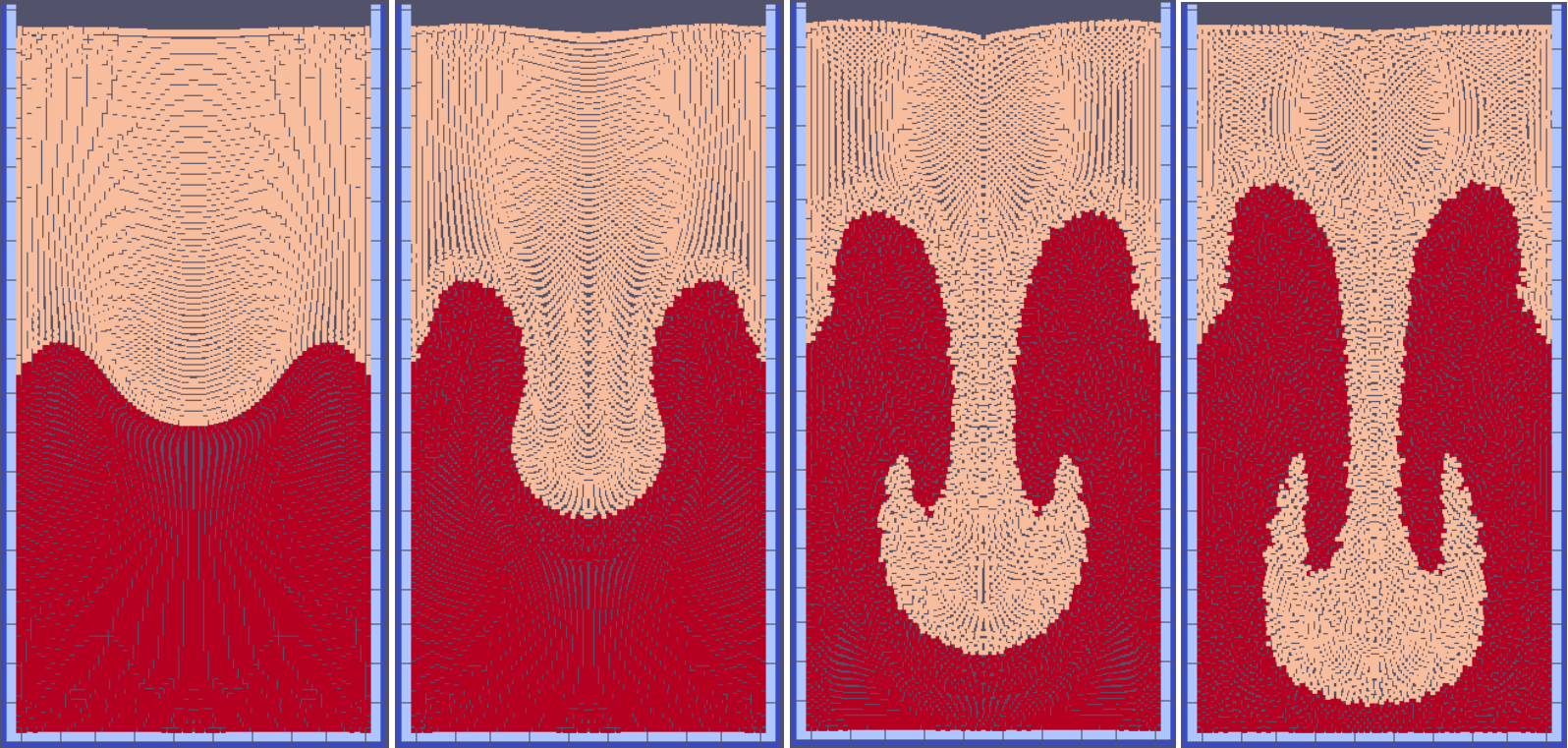}
  \caption{Rayleigh-Taylor instability generated by a multiphase interaction developed in this work}
  \label{fig:rt-insta-our}
\end{subfigure}
\begin{subfigure}{\linewidth}
  \centering
  \includegraphics[width=.8\linewidth]{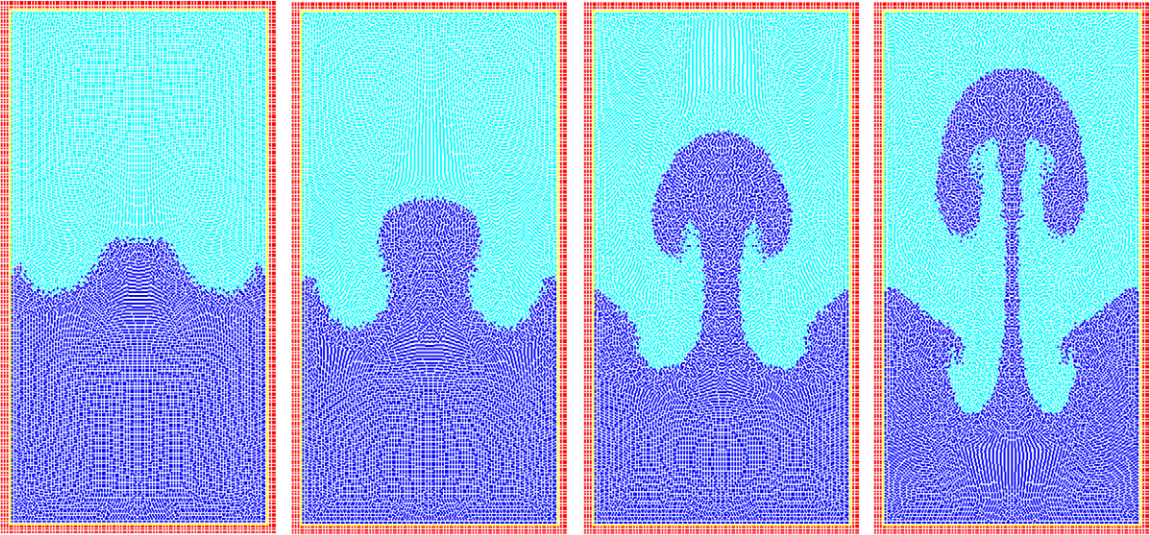}
  \caption{Rayleigh-Taylor instability by a basic multiphase MPS. Extracted from \cite{shakibaeinia2012mps}}
  \label{fig:rt-insta-shaki}
\end{subfigure}

\caption{Rayleigh-Taylor instability in almost similar scenarios}
\label{fig:rt-insta}
\end{figure}

The proposed technique generally shows good agreement regarding the Rayleigh-Taylor instability when compared to another particle-based solution.

\subsection{Oil spill}

The Oil spill test makes possible a qualitative analysis and a unique comparison to validate the multiphase model. \cite{duan2017a} provides a replicable test of a continuous oil spill due to a damaged tank. 

The average particle distance is $4\times 10^{-3}~m$, the water and oil's kinematic viscosity $\nu_{water}$ and $\nu_{oil}$ are, respectively, $10^{-6}~m^2/s$ and $5 \times 10^{-5}~m^2/s$ and configuration here, relative to \autoref{fig:gui}, differs in fields (3), which is Oil Spill, (4), Weakly compressible, (11), which is $897.0 Kg/m^3$ and (12), which is $5 \times 10^{-5}~m^2/s$. This test has a total of $152,779$ particles.

This leakage simulation is qualitatively compared to experimental results and to \cite{duan2017a}, which is a fully incompressible multiphase MPS. \autoref{fig:oilspilling} shows five time instants of the experiments and the simulations. In each time instants, there is: an image of the experiment; the simulation by \cite{duan2017a}; and the generated simulation, in that order.

\begin{figure}[ht]
\centering
\begin{subfigure}{0.49\linewidth}
  \centering
  \includegraphics[width=0.99\linewidth]{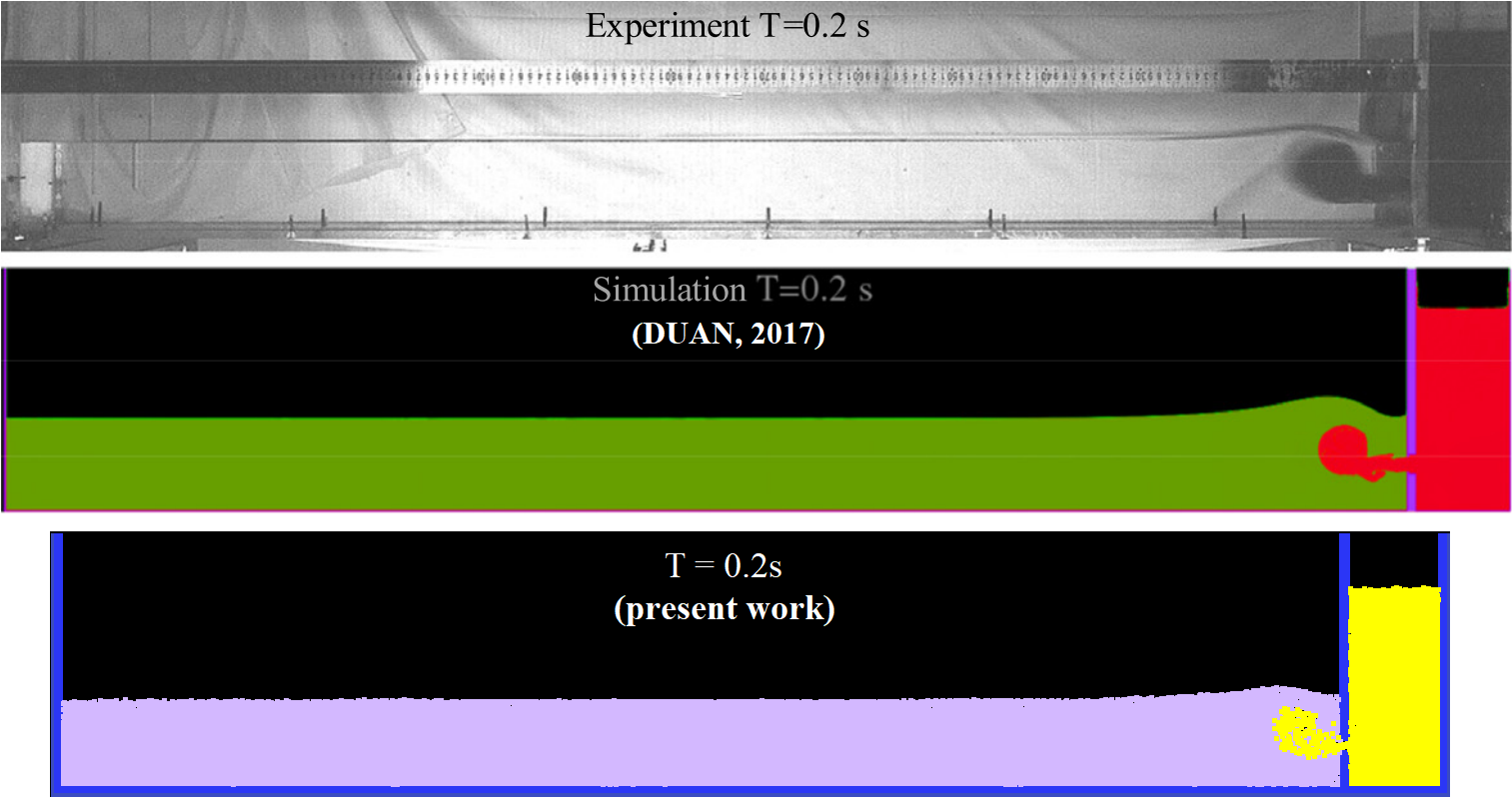}
  \caption{$t = 0.2~s$}
  \label{fig:oil0.2}
\end{subfigure}
\begin{subfigure}{0.49\linewidth}
  \centering
  \includegraphics[width=0.99\linewidth]{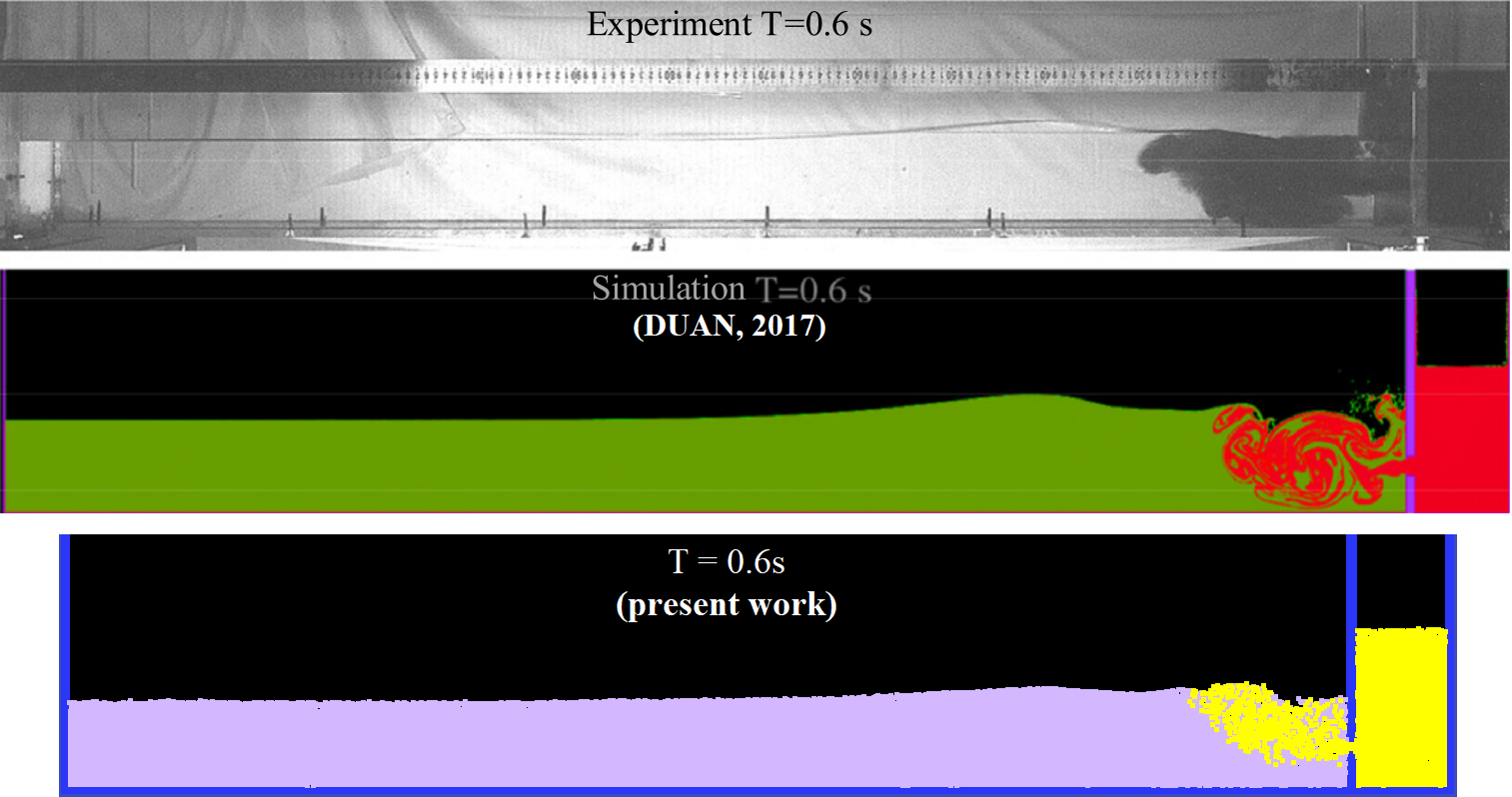}
  \caption{$t = 0.6~s$}
  \label{fig:oil0.6}
\end{subfigure}

\begin{subfigure}{0.49\linewidth}
  \centering
  \includegraphics[width=0.99\linewidth]{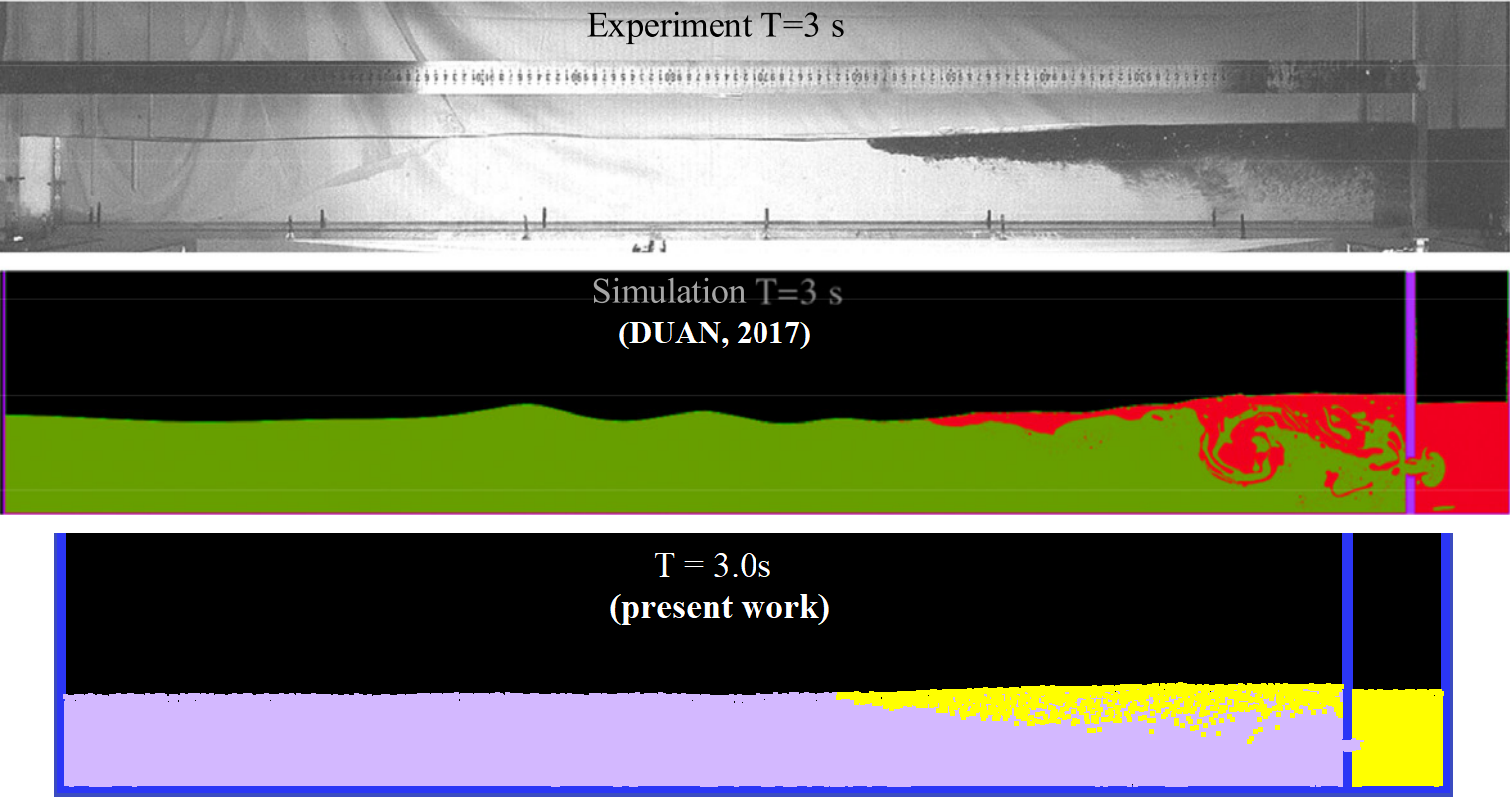}
  \caption{$t = 3~s$}
  \label{fig:oil3}
\end{subfigure}
\begin{subfigure}{0.49\linewidth}
  \centering
  \includegraphics[width=0.99\linewidth]{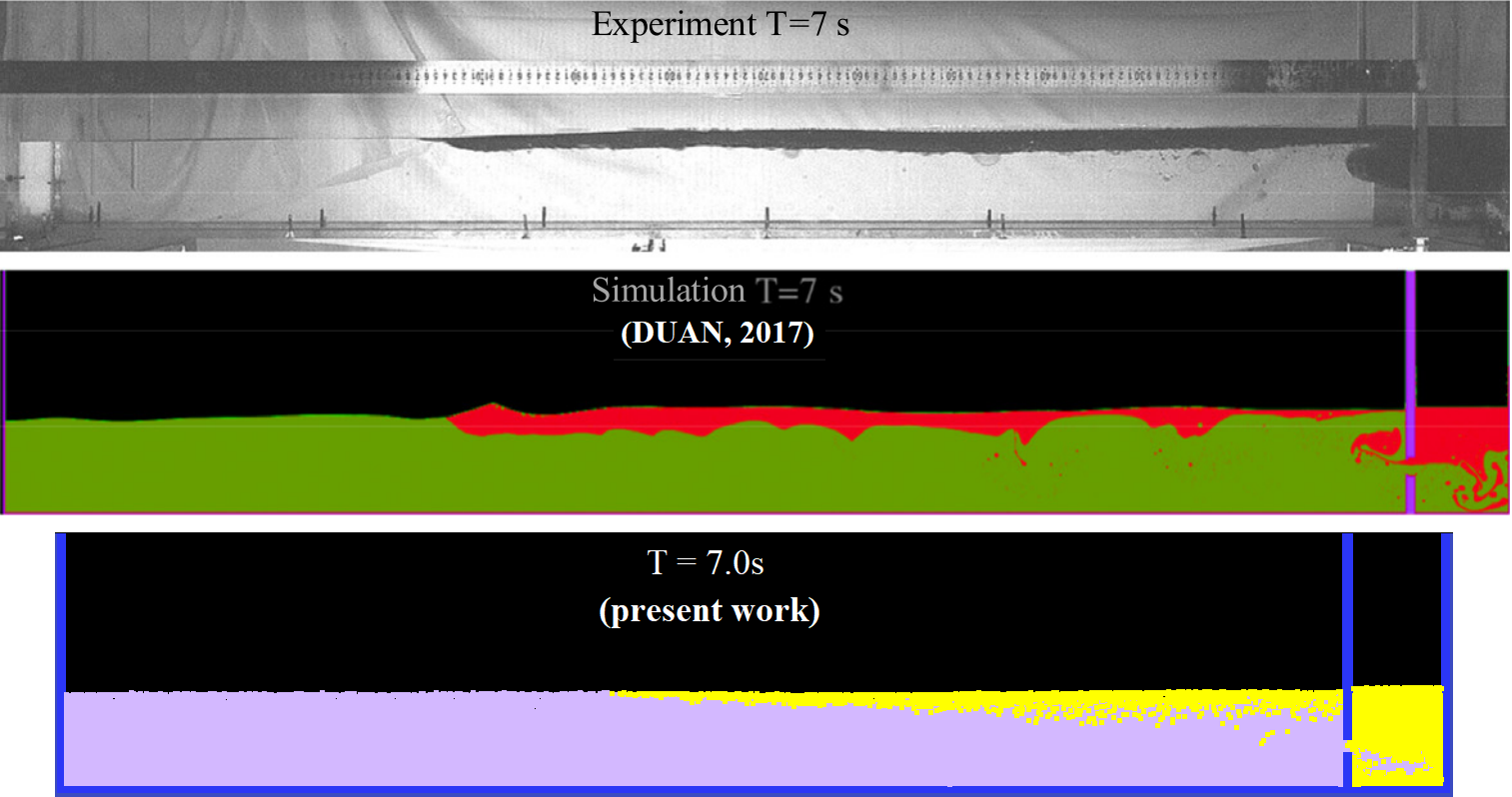}
  \caption{$t = 7~s$}
  \label{fig:oil7}
\end{subfigure}

\caption{Comparison between experiment, FS-MMPS \cite{duan2017a} and the developed system at different moments of the oil spill}
\label{fig:oilspilling}
\end{figure}

The authors in \cite{duan2017a} claim that FS-MMPS is quite precise and accurate. On the other hand, it is possible to observe that the use of a weakly compressible approach of the combined method developed here can better predict certain situations, such as the more straight and even flow of the oil, in yellow, over the denser fluid, mainly in \autoref{fig:oil3} and \autoref{fig:oil7}. In a general manner, the oil profile and the waves generated by it are in reasonable agreement with the generated by the experiment, with fewer computations than the FS-MMPS, since, in this case, a WC model is used, which significantly diminishes the pressure calculation complexity.

\subsection{Pressure field}

It is essential to show how the previously presented numerical improvements impact the pressure field of the fluid. This test makes use of the traditional dam break scenario described in \autoref{sec:dambreak}. \autoref{fig:pressurefield} shows three time instants of the same simulation. The left side refers to the simulation using the standard MPS, and the right side, the simulation with the numerical improvements described in \autoref{subsec:enhancements}.

\begin{figure}
\centering
\includegraphics[width=0.99\linewidth]{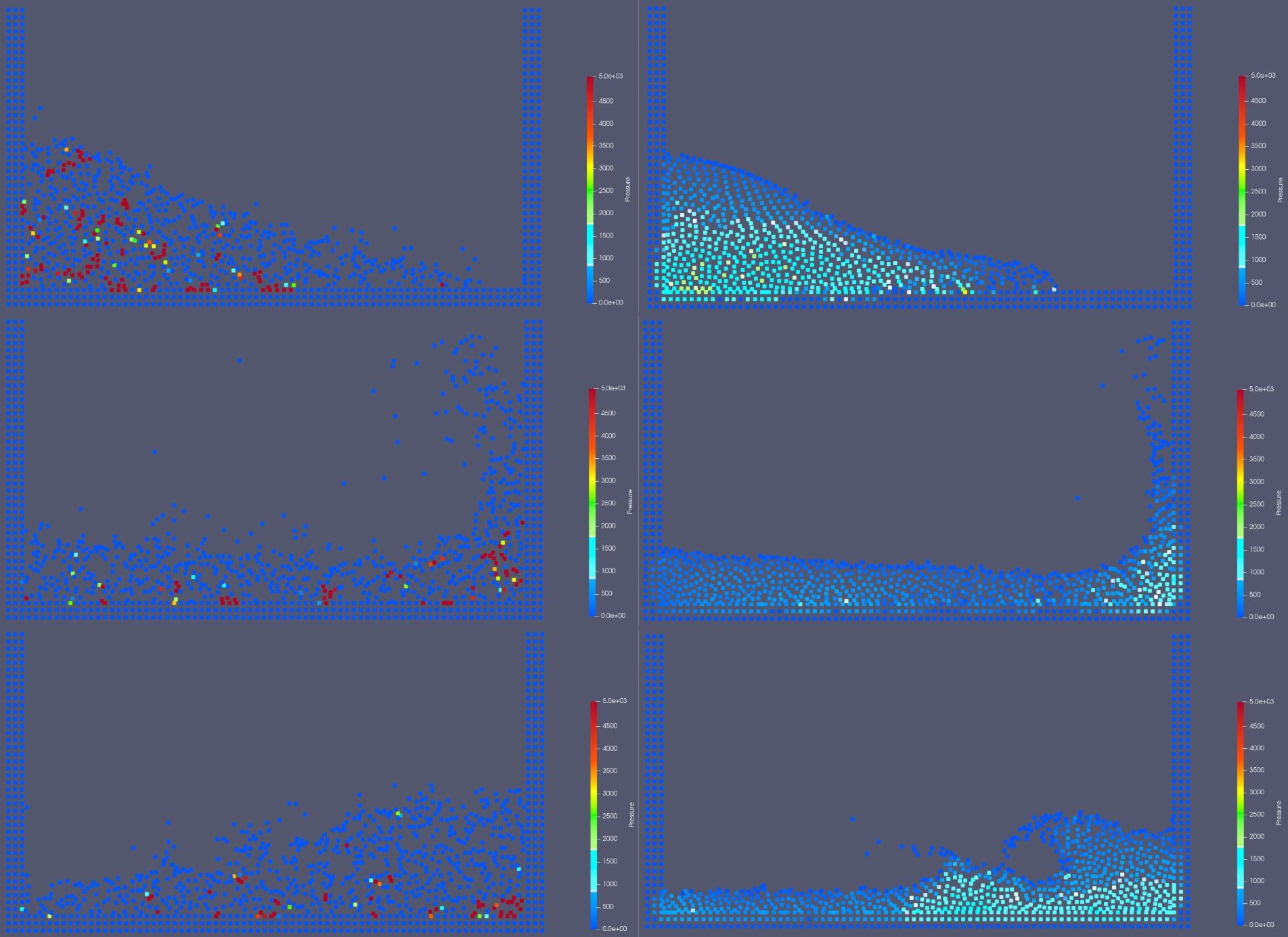}
\caption{\label{fig:pressurefield} Dam break simulation. Left side: standard MPS; Right side: modified MPS with numerical improvements described in \autoref{subsec:enhancements}. Time instants referring 0.25~s, 0.5~s and 0.875~s into the simulation.}
\end{figure}

It is apparent in \autoref{fig:pressurefield} that the numerical improvements significantly diminish the spurious pressure oscillations and can provide a more stable pressure field during the simulation, even after the fluid suffers large deformations.

\subsection{Computational performance analysis}

It is noticeable that switching between the incompressibility models - weakly compressible (WC) or fully incompressible (FI) - had a material impact on computational performance, both in memory usage and in runtime performance. This discrepancy happens because the FI approach requires the solution of a linear system of order $N \times N_{neigh}$, to ensure incompressibility and, thus, obtain more precise values for the pressures. Storing it in memory and solving it are both costly steps for the algorithm. The chosen incompressibility model influences the most on the runtime, regardless of the selected combination of enhancements. 

To evaluate the computational performance, this section details the memory usage and the runtime profile of two main approaches (WC and FI) combined with three different types of execution: completely sequential; parallelized through OpenMP; parallelized through CUDA. The entire runtime for each test size was measured to calculate the speedups. Finally, it is possible to calculate each version's frame rate by running a set of tests in which they vary in the number of particles.

\subsubsection{Memory usage} \label{sec:mem}

The Performance and Diagnostics tool of Visual Studio 2019 \cite{msvs} provides a detailed report of the sequential and OpenMP executions to evaluate CPU memory usage in all versions. The NVIDIA Visual profiler \cite{nvprof} provided the majority of information regarding the GPU, such as the whole simulation runtime and time spent in each CUDA kernel during the execution. As for the GPU memory usage, GPU-Z \cite{gpuz} provided the minimum and maximum occupancy throughout the simulation runtime.

\autoref{relmemwc} shows the memory usage of the WC approach of the MPS for a standard 2D dam break test, like the one presented in \autoref{sec:dambreak}. For this case, there is a total of $1,054,122$ particles.

\begin{table}
\centering
\caption{Occupancy in the CPU and GPU memories of different executions of the WC-MPS for 1 million particles}
\label{relmemwc}
\begin{tabular}{cccccc}
        & \vline & Sequential & OpenMP & CUDA \\ \hline
CPU Mem. usage & \vline & $4.7~\%$ & $4.7~\%$ & $4.4~\%$              \\
GPU Mem. usage & \vline & 0~\% & 0~\%  & $16.4~\%$              
\end{tabular}
\end{table}

\autoref{relmemfi} shows the memory usage relative to the FI approach for the same test described above and in \autoref{sec:dambreak}. One million particles can be considered a high amount of particles for a FI approach running on GPU since a linear system has to be fully loaded. This number of particles is an improvement compared to previous works such as \cite{vieira-e-silva2017improved, vieira-e-silva2018improved}.

\begin{table}
\centering
\caption{Occupancy in the CPU and GPU memories of different executions of the FI-MPS for 1 million particles}
\label{relmemfi}
\begin{tabular}{cccccc}
        & \vline & Sequential & OpenMP & CUDA \\ \hline
CPU Mem. usage & \vline & $20.3~\%$ & $20.3~\%$ & $20.0~\%$              \\
GPU Mem. usage & \vline & 0~\% & 0~\%  & $60.0~\%$              
\end{tabular}
\end{table}

Note that, in both approaches, the CUDA versions require not only the GPU memory but also CPU memory, which is also smaller compared to OpenMP and the fully sequential executions. This difference may happen since some information must exist in host memory (CPU) and in device memory (GPU) to transfer particle input information from the CPU to the GPU.

\subsubsection{Speedup}\label{subsec:speed}

Due to the scalability of the GPU-accelerated neighbourhood search strategy, the system was able to reach promising speedup values. \autoref{fig:speedup_wc} shows that this is especially true in more massive simulations. The OpenMP version did not achieve considerable speedups since its neighbourhood calculation is not optimally parallelized.

\begin{figure}[ht]
\centering
\includegraphics[width=0.75\linewidth]{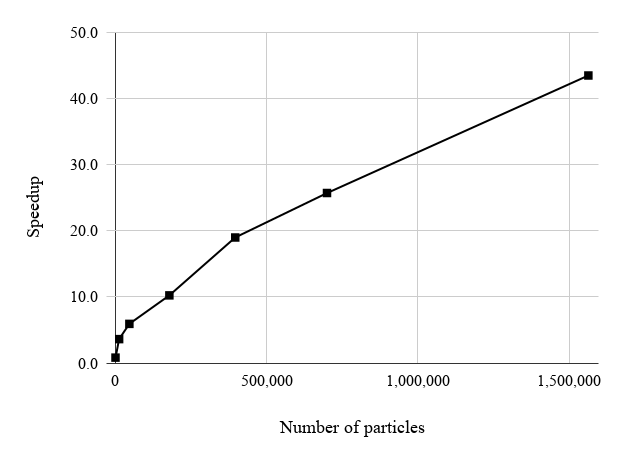}
\caption{\label{fig:speedup_wc} Speedup caused by GPU parallelization versus total number of particles in the simulation}
\end{figure}

\autoref{fig:speedup_wc} shows that a case with $1,563,658$ particles can achieved a massive speedup of $43.5$ times. The speedup curve shows that simulations with more particles could yield even higher speedup rates. On the other hand, for a case with $1,022$ particles, the speedup rate obtained was $0.8$ times, in other words, running this scenario in the GPU is slower than running it in sequentially in the CPU. That shows it is not efficient to employ parallel programming for simulations with a small number of particles. One of the causes for this is that the data transfer between the host and device reduces the speedup significantly.

\subsubsection{System limits}\label{subsec:limits}

Data related to performance and memory limitations are presented here, such as frame rate achieved for different numbers of particles and the maximum number of particles in a simulation. 

\autoref{fig:fps} shows how many frames per second (FPS) the GPU-accelerated application achieves, given the total number of particles in the simulation. The blue frame rate curve represents a regular execution where the neighbourhood is updated every step, prioritizing a more accurate simulation. The red frame rate curve represents an execution where the list of neighbours is only updated every four steps during the simulation, aiming for high computational performance.

\begin{figure}[ht]
\centering
\includegraphics[width=0.8\linewidth]{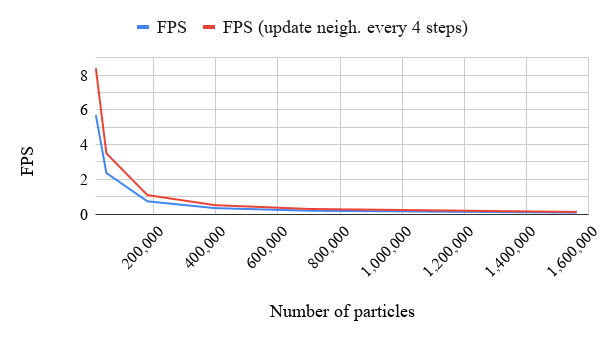}
\caption{\label{fig:fps} Frame rate of the simulation given its total number of particles.}
\end{figure}

Hence, in a case that computational performance is the priority, \autoref{fig:fps} shows that it is possible to simulate a scenario with approximately $200,000$ particles at one frame per second. It also displays a case with $46,768$ particles run at approximately 4 fps. Both of them are already considered interactive frame rates \cite{wald2004an,overby2002interactive}.

Regarding the simulation size limit, it is noteworthy that the available memory plays a fundamental role. The GPU used in this work provides 11 GB of video memory and allows for a simulation with $1,563,658$ before the application runs out of memory. The largest simulation size tested for the fully sequential version when adopting the fully incompressible approach had a total of $3,677,442$ particles. The CPU process of the latter used 25 of the 32 GB available of RAM. When adopting the weakly compressible model, the total number of particles could reach $16,657,962$ by using 26 GB of the available RAM. The last two simulation loads show how expensive it is in terms of memory, just loading the PPE into the RAM, not to mention the computational load required to solve it.

\section{Conclusions}
\label{chap:conclusoecvcs}

As previously discussed, the study of fluid flow simulation is of great importance in mitigating the consequences of environmental disasters and accidents. It has applications in a wide range of engineering problems, computer graphics, and virtual and augmented reality software. Meshless methods like the MPS are a great alternative to deal with large deformations and free-surface flow, situations where the traditional mesh-based approaches usually perform inefficiently.

Throughout the development of his work, it became clear that the community is continually improving the MPS, both in numerical performance and computation efficiency}, despite suffering from a few pressure instability problems. The literature shows its usability in a large number of scenarios. The considerable amount of referenced works also shows the complexity of this task, the importance of the method, and the great potential to simulate, increasingly more realistically, fluid flows. 

MPS optimization is moderately complex since it is used to replicate real phenomena more reliably, when taking into consideration other meshless approaches. Some works enhance its computational efficiency with acceleration structures without losing the precision it offers \cite{hori2011gpu} \cite{zhu2011implementation} \cite{vieira-e-silva2017improved} \cite{vieira-e-silva2018improved}. In contrast, other works prefer, despite the precision loss, replace the performance and memory bottleneck of the method, the solution of a PPE, with an equation of state to solve the pressures \cite{shakibaeinia2010weakly} \cite{tayebi2015development} and then, parallelize it through GPU \cite{taniguchi2014explicit} \cite{fernandes2013phd}. Another factor to consider is hardware development since manufacturers are continuously building more powerful GPUs, which directly influence computational performance.

The literature also shows that works usually accelerate the standard version of the MPS \cite{koshizuka1996moving} or the standard weakly compressible version \cite{shakibaeinia2010weakly}\cite{tayebi2015development}, with few numerical improvements to the calculation. This work provides a wide variety of models, improvements, and approaches to the MPS technique, which are entirely parallelized, through OpenMP and CUDA, integrated into a single solution. This integration enables a fine-tuning of the system, allowing setups concerned with high precision, fully incompressible fluid flow, and GPU-accelerated multiphase WC fluid simulations. This solution is open-source under the GPLv3 license.

Regarding the numerical improvements, the techniques proposed by \cite{gotoh2013advanced} and \cite{shakibaeinia2012mps} were combined and extended, which implicated in enhancements in simulation coherence, as presented in the Dam break, Oil spill, and Water drop tests.
The developed method shows compatibility with recent works in the Oil Spill test, by qualitatively comparing it with \cite{duan2017a}, which shows the numerical advances achieved. The GPU-acceleration provides speedups ranging from 3 to 43 times, depending on the total number of particles in the simulation. That permits a simulation with approximately $200,000$ particles to run at one frame per second, and a test case with $46,768$ particles to achieve nearly 4 fps, which both are already considered interactive rates \cite{wald2004an,overby2002interactive}.  

\subsection{Future Work}

There are exciting possibilities for future developments of this work. Inevitably, refinements in every part of the code will lead to a more optimized version, which is the path to a notable real-time simulation.

A possibility that would enhance even further computational performance is to improve the GPU implementation of the FI-MPS to be able to run locally on multiple GPU or in a GPU cluster, which could raise the speedup values to new levels. \cite{taniguchi2014explicit} shows the performance of WC-MPS aided by a GPU cluster; fully incompressible versions of the MPS could benefit significantly from such structures. \cite{fernandes2015domain} presents a domain decomposition strategy for a parallelized MPS for running in a cluster of computers. That enables massive simulations since the simulation domain can be loaded and solved separately in the PC's memories and then integrated back.

Another possibility is to improve the OpenMP parallelization of the neighbourhood search function. Since that function is based on its sequential version, it could not reach decent speedup values. However, with an optimal CPU parallelization of that step, the OpenMP acceleration could reach new levels.

Still on the acceleration task, since solving a PPE is costly, an alternative to the code parallelization could be the usage of neural networks. They would learn the usual results of commonly yielded linear systems by fully incompressible fluid simulations. The work of \cite{tompson2016accelerating} shows promising fluid simulations using this approach. 

The addition of models that allow the interaction between fluids and solids, such as floating bodies, deformable bodies, and viscoelastic fluid, would drastically increase the number of application possibilities.



\section*{Acknowledgements}

The authors would like to thank Mr. Deep Tavker for authorizing the use of part of his work to help develop the neighbourhood search strategy adopted here.
\bibliographystyle{unsrt}
\bibliography{refs}

\begin{thebibliography}{10}

\bibitem{cleary2006novel}
PW~Cleary, M~Prakash, and J~Ha.
\newblock Novel applications of smoothed particle hydrodynamics ({SPH}) in
  metal forming.
\newblock {\em Journal of materials processing technology}, 177(1):41--48,
  2006.

\bibitem{muller2003particle}
Matthias M\"{u}ller, David Charypar, and Markus Gross.
\newblock Particle-based fluid simulation for interactive applications.
\newblock In {\em Proceedings of the 2003 ACM SIGGRAPH/Eurographics Symposium
  on Computer Animation}, SCA '03, pages 154--159, Aire-la-Ville, Switzerland,
  Switzerland, 2003. Eurographics Association.

\bibitem{daenzer2007real}
Stefan Daenzer, Kevin Montgomery, R{\"u}diger Dillmann, and Roland
  Unterhinninghofen.
\newblock Real-time smoke and bleeding simulation in virtual surgery.
\newblock In {\em MMVR}, pages 94--99, 2007.

\bibitem{bridson2015fluid}
Robert Bridson.
\newblock {\em Fluid simulation for computer graphics}.
\newblock CRC press, 2015.

\bibitem{belytschko1996meshless}
Ted Belytschko, Yury Krongauz, Daniel Organ, Mark Fleming, and Petr Krysl.
\newblock Meshless methods: an overview and recent developments.
\newblock {\em Computer methods in applied mechanics and engineering},
  139(1):3--47, 1996.

\bibitem{johnson1999advanced}
Andrew~A Johnson and Tayfun~E Tezduyar.
\newblock Advanced mesh generation and update methods for 3d flow simulations.
\newblock {\em Computational Mechanics}, 23(2):130--143, 1999.

\bibitem{frey2005anisotropic}
Pascal-Jean Frey and Fr{\'e}d{\'e}ric Alauzet.
\newblock Anisotropic mesh adaptation for {CFD} computations.
\newblock {\em Computer methods in applied mechanics and engineering},
  194(48):5068--5082, 2005.

\bibitem{koshizuka1996moving}
S~Koshizuka and Y~Oka.
\newblock Moving-particle semi-implicit method for fragmentation of
  incompressible fluid.
\newblock {\em Nuclear science and engineering}, 123(3):421--434, 1996.

\bibitem{gingold1977smoothed}
Robert~A Gingold and Joseph~J Monaghan.
\newblock Smoothed particle hydrodynamics: theory and application to
  non-spherical stars.
\newblock {\em Monthly notices of the royal astronomical society},
  181(3):375--389, 1977.

\bibitem{lucy1977numerical}
Leon~B Lucy.
\newblock A numerical approach to the testing of the fission hypothesis.
\newblock {\em The astronomical journal}, 82:1013--1024, 1977.

\bibitem{gotoh2016current}
Hitoshi Gotoh and Abbas Khayyer.
\newblock Current achievements and future perspectives for projection-based
  particle methods with applications in ocean engineering.
\newblock {\em Journal of Ocean Engineering and Marine Energy}, 2(3):251--278,
  Aug 2016.

\bibitem{gotoh1999lagrangian}
Hitoshi Gotoh and Tetsuo Sakai.
\newblock Lagrangian simulation of breaking waves using particle method.
\newblock {\em Coastal Engineering Journal}, 41(03n04):303--326, 1999.

\bibitem{khayyer2008}
Abbas Khayyer.
\newblock {\em Improved particle methods by refined models for free-surface
  fluid flows}.
\newblock PhD thesis, Kyoto University, 2008.

\bibitem{gotoh2005lagrangian}
Hitoshi Gotoh, Hiroyuki Ikari, Tetsu Memita, and Tetsuo Sakai.
\newblock Lagrangian particle method for simulation of wave overtopping on a
  vertical seawall.
\newblock {\em Coastal Engineering Journal}, 47(02n03):157--181, 2005.

\bibitem{khayyer2009modified}
Abbas Khayyer and Hitoshi Gotoh.
\newblock Modified moving particle semi-implicit methods for the prediction of
  2d wave impact pressure.
\newblock {\em Coastal Engineering}, 56(4):419--440, 2009.

\bibitem{lee2011step}
Byung-Hyuk Lee, Jong-Chun Park, Moo-Hyun Kim, and Sung-Chul Hwang.
\newblock Step-by-step improvement of {MPS} method in simulating violent
  free-surface motions and impact-loads.
\newblock {\em Computer methods in applied mechanics and engineering},
  200(9):1113--1125, 2011.

\bibitem{shibata2007numerical}
Kazuya Shibata and Seiichi Koshizuka.
\newblock Numerical analysis of shipping water impact on a deck using a
  particle method.
\newblock {\em Ocean Engineering}, 34(3):585--593, 2007.

\bibitem{gotoh2006key}
Hitoshi Gotoh and Tetsuo Sakai.
\newblock Key issues in the particle method for computation of wave breaking.
\newblock {\em Coastal Engineering}, 53(2):171 -- 179, 2006.
\newblock Coastal Hydrodynamics and Morphodynamics.

\bibitem{fu2015investigation}
Lei Fu and Yee-Chung Jin.
\newblock Investigation of non-deformable and deformable landslides using
  meshfree method.
\newblock {\em Ocean Engineering}, 109:192 -- 206, 2015.

\bibitem{shibata2012lagrangian}
Kazuya Shibata, Seiichi Koshizuka, Mikio Sakai, and Katsuji Tanizawa.
\newblock Lagrangian simulations of ship-wave interactions in rough seas.
\newblock {\em Ocean Engineering}, 42:13 -- 25, 2012.

\bibitem{hwang2014development}
Sung-Chul Hwang, Abbas Khayyer, Hitoshi Gotoh, and Jong-Chun Park.
\newblock Development of a fully lagrangian {MPS}-based coupled method for
  simulation of fluid-structure interaction problems.
\newblock {\em Journal of Fluids and Structures}, 50:497 -- 511, 2014.

\bibitem{zhu2011implementation}
XiaoSong Zhu, Liang Cheng, Lin Lu, and Bin Teng.
\newblock Implementation of the moving particle semi-implicit method on {GPU}.
\newblock {\em SCIENCE CHINA Physics, Mechanics \& Astronomy}, 54(3):523--532,
  2011.

\bibitem{hori2011gpu}
Chiemi Hori, Hitoshi Gotoh, Hiroyuki Ikari, and Abbas Khayyer.
\newblock {GPU}-acceleration for moving particle semi-implicit method.
\newblock {\em Computers \& Fluids}, 51(1):174--183, 2011.

\bibitem{taniguchi2014explicit}
D~Taniguchi, LM~Sato, and LY~Cheng.
\newblock Explicit moving particle simulation method on {GPU} clusters.
\newblock {\em Blucher Mech. Eng. Proc. 1}, 1:1155, 2014.

\bibitem{dagum1998openmp}
L.~Dagum and R.~Menon.
\newblock Openmp: an industry standard api for shared-memory programming.
\newblock {\em IEEE Computational Science and Engineering}, 5(1):46--55, Jan
  1998.

\bibitem{cudatoolkit}
NVIDIA.
\newblock {CUDA Zone | NVIDIA Developer}, 2007.
\newblock Accessed: 2016-01-09.

\bibitem{chen2013frontiers}
G.~Chen, Y.~Onishi, L.~Zheng, and T.~Sasaki.
\newblock {\em Frontiers of Discontinuous Numerical Methods and Practical
  Simulations in Engineering and Disaster Prevention}.
\newblock Taylor \& Francis Group, London, 8 2013.

\bibitem{almeida2016meshless}
M.~Almeida, C.~Brito, A.~L.~B. {Vieira-e-Silva}, V.~Teichrieb, and J.~M.
  Barbosa.
\newblock Meshless methods.
\newblock In G.~Assi, H.~Brinati, M.~de~Conti, and M.~Szajnbok, editors, {\em
  Applied Topics in Marine Hydrodynamics}, chapter~8, pages 8.1--8.38. Escola
  Polit{\'e}cnica da Universidade de S{\~a}o Paulo (ISBN 978-85-86686-89-4),
  S{\~a}o Paulo, 2016.

\bibitem{mpsryujin}
{Fuji Technical Research Inc.}
\newblock {MPS-RYUJIN}, 2013.
\newblock Accessed: 2018-04-23.

\bibitem{particleworks}
{Prometech Software}.
\newblock Particleworks, 2014.
\newblock Accessed: 2018-04-23.

\bibitem{mpars}
Ahmad Shakibaeinia.
\newblock {MPARS - Mesh-free Particle Simulator}, 2012.
\newblock Accessed: 2018-04-09.

\bibitem{crespo2015dualsphysics}
Alejandro~JC Crespo, Jos{\'e}~M Dom{\'\i}nguez, Benedict~D Rogers, Moncho
  G{\'o}mez-Gesteira, S~Longshaw, R~Canelas, Renato Vacondio, A~Barreiro, and
  O~Garc{\'\i}a-Feal.
\newblock Dualsphysics: Open-source parallel {CFD} solver based on smoothed
  particle hydrodynamics ({SPH}).
\newblock {\em Computer Physics Communications}, 187:204--216, 2015.

\bibitem{gpusph}
Alexis H\'erault, Robert~A. Dalrymple, Billy Edge, Giuseppe Bilotta, and
  Agn\`es Leroy.
\newblock {GPUSPH}.
\newblock Accessed: 2019-05-21.

\bibitem{kondoincompressible}
Masahiro Kondo, Kentaro Suto, Mikio Sakai, and Seiichi Koshizuka.
\newblock Incompressible free surface flow analysis using moving particle
  semi-implicit method.
\newblock {\em Joint International Workshop: Nuclear Technology and Society –
  Needs for Next Generation, Berkeley, California, January 6-8, 2008, Berkeley
  Faculty Club, UC Berkeley Campus}, 2008.

\bibitem{khayyer2010higher}
Abbas Khayyer and Hitoshi Gotoh.
\newblock A higher order laplacian model for enhancement and stabilization of
  pressure calculation by the {MPS} method.
\newblock {\em Applied Ocean Research}, 32(1):124--131, 2010.

\bibitem{khayyer2011enhancement}
Abbas Khayyer and Hitoshi Gotoh.
\newblock Enhancement of stability and accuracy of the moving particle
  semi-implicit method.
\newblock {\em Journal of Computational Physics}, 230(8):3093--3118, 2011.

\bibitem{MONAGHAN2000290}
J.J. Monaghan.
\newblock {SPH} without a tensile instability.
\newblock {\em Journal of Computational Physics}, 159(2):290 -- 311, 2000.

\bibitem{gotoh2018on}
Hitoshi Gotoh and Abbas Khayyer.
\newblock On the state-of-the-art of particle methods for coastal and ocean
  engineering.
\newblock {\em Coastal Engineering Journal}, 0(0):1--25, 2018.

\bibitem{tayebi2015development}
Ali Tayebi and Yee chung Jin.
\newblock Development of moving particle explicit (mpe) method for
  incompressible flows.
\newblock {\em Computers \& Fluids}, 117:1 -- 10, 2015.

\bibitem{duan2017a}
Guangtao Duan, Bin Chen, Ximin Zhang, and Yechun Wang.
\newblock A multiphase {MPS} solver for modeling multi-fluid interaction with
  free surface and its application in oil spill.
\newblock {\em Computer Methods in Applied Mechanics and Engineering}, 320:133
  -- 161, 2017.

\bibitem{harada2017numerical}
Eiji Harada, Hitoshi Gotoh, Hiroyuki Ikari, and Abbas Khayyer.
\newblock Numerical simulation for sediment transport using {MPS}-{DEM}
  coupling model.
\newblock {\em Advances in Water Resources}, 2017.

\bibitem{vieira-e-silva2018improved}
Andr\'e Luiz~Buarque {Vieira-e-Silva}, Mozart W.~S. Almeida, Caio Brito, and
  Veronica Teichrieb.
\newblock Improved \uppercase{MPS} method and its variations for simulating
  incompressible fluids on \uppercase{GPU}.
\newblock {\em Journal on 3D Interactive Systems}, 9(2), 2018.

\bibitem{shakibaeinia2010weakly}
Ahmad Shakibaeinia and Yee‐Chung Jin.
\newblock A weakly compressible {MPS} method for modeling of open‐boundary
  free‐surface flow.
\newblock {\em International Journal for Numerical Methods in Fluids},
  63(10):1208--1232, 2010.

\bibitem{shakibaeinia2012mps}
Ahmad Shakibaeinia and Yee-Chung Jin.
\newblock {MPS} mesh-free particle method for multiphase flows.
\newblock {\em Computer Methods in Applied Mechanics and Engineering},
  229-232:13 -- 26, 2012.

\bibitem{koshizuka1998numerical}
Seiichi Koshizuka, Atsushi Nobe, and Yoshiaki Oka.
\newblock Numerical analysis of breaking waves using the moving particle
  semi-implicit method.
\newblock {\em International Journal for Numerical Methods in Fluids},
  26(7):751--769, 1998.

\bibitem{batchelor1970k}
G~Batchelor.
\newblock K. 1967 an introduction to fluid dynamics, 1970.

\bibitem{monaghan1994simulating}
Joe~J Monaghan.
\newblock Simulating free surface flows with {SPH}.
\newblock {\em Journal of computational physics}, 110(2):399--406, 1994.

\bibitem{monaghan2005smoothed}
J~J Monaghan.
\newblock Smoothed particle hydrodynamics.
\newblock {\em Reports on Progress in Physics}, 68(8):1703, 2005.

\bibitem{batchelor2000introduction}
Cx~K Batchelor and GK~Batchelor.
\newblock {\em An introduction to fluid dynamics}.
\newblock Cambridge university press, 2000.

\bibitem{shao2005turbulence}
Songdong Shao and Hitoshi Gotoh.
\newblock Turbulence particle models for tracking free surfaces.
\newblock {\em Journal of Hydraulic Research}, 43(3):276--289, 2005.

\bibitem{smagorinsky1963general}
J.~Smagorinsky.
\newblock General circulation experiments with the primitive equations.
\newblock {\em Monthly Weather Review}, 91(3):99--164, 1963.

\bibitem{rogallo1984numerical}
Robert~S Rogallo and Parviz Moin.
\newblock Numerical simulation of turbulent flows.
\newblock {\em Annual Review of Fluid Mechanics}, 16(1):99--137, 1984.

\bibitem{khayyer2008development}
Abbas Khayyer and Hitoshi Gotoh.
\newblock Development of {CMPS} method for accurate water-surface tracking in
  breaking waves.
\newblock {\em Coastal Engineering Journal}, 50(02):179--207, 2008.

\bibitem{khayyer20123d}
Abbas Khayyer and Hitoshi Gotoh.
\newblock A 3d higher order laplacian model for enhancement and stabilization
  of pressure calculation in 3d {MPS}-based simulations.
\newblock {\em Applied Ocean Research}, 37:120--126, 2012.

\bibitem{gotoh2013advanced}
H~Gotoh.
\newblock Advanced particle methods for accurate and stable computation of
  fluid flows.
\newblock {\em Frontiers of Discontinuous Numerical Methods and Practical
  Simulations in Engineering and Disaster Prevention}, page 113, 2013.

\bibitem{platt2002introducing}
David~S. Platt.
\newblock {\em Introducing Microsoft .Net, Second Edition}.
\newblock Microsoft Press, Redmond, WA, USA, 2nd edition, 2002.

\bibitem{inteli74790}
Intel.
\newblock Intel processor i7 4790 specifications, 2013.
\newblock Accessed: 2018-04-22.

\bibitem{tavker2018parallel}
Deep Tavker.
\newblock {Parallel Neighbour Search Implementation}.
\newblock \url{https://github.com/deeptavker/Parallel-Neighbour-Search}, 2018.

\bibitem{bonet1999variational}
J~Bonet and T-SL Lok.
\newblock Variational and momentum preservation aspects of smooth particle
  hydrodynamic formulations.
\newblock {\em Computer Methods in applied mechanics and engineering},
  180(1):97--115, 1999.

\bibitem{hirt1981volume}
C.W Hirt and B.D Nichols.
\newblock Volume of fluid (vof) method for the dynamics of free boundaries.
\newblock {\em Journal of Computational Physics}, 39(1):201 -- 225, 1981.

\bibitem{koshizuka1995particle}
Seichii Koshizuka.
\newblock A particle method for incompressible viscous flow with fluid
  fragmentation.
\newblock {\em Comput. Fluid Dynamics J.}, 4:29--46, 1995.

\bibitem{martin1952experimental}
J.~C. Martin, W.~J. Moyce, J.~C. Martin, W.~J. Moyce, W.~G. Penney, F.~R. S.,
  A.~T. Price, and C.~K. Thornhill.
\newblock Part iv. an experimental study of the collapse of liquid columns on a
  rigid horizontal plane.
\newblock {\em Philosophical Transactions of the Royal Society of London A:
  Mathematical, Physical and Engineering Sciences}, 244(882):312--324, 1952.

\bibitem{youngs1984numerical}
David~L. Youngs.
\newblock Numerical simulation of turbulent mixing by rayleigh-taylor
  instability.
\newblock {\em Physica D: Nonlinear Phenomena}, 12(1):32 -- 44, 1984.

\bibitem{msvs}
Microsoft.
\newblock {Visual Studio Enterprise 2015}, 2018.
\newblock Accessed: 2018-05-01.

\bibitem{nvprof}
NVIDIA.
\newblock {NVIDIA Visual Profiler}, 2018.
\newblock Accessed: 2018-05-01.

\bibitem{gpuz}
TechPowerUp.
\newblock {GPU-Z Video card GPU Information Utility}, 2018.
\newblock Accessed: 2018-05-01.

\bibitem{vieira-e-silva2017improved}
A.~L.~B. {Vieira-e-Silva}, M.~W.~S. Almeida, C.~Brito, and V.~Teichrieb.
\newblock Improved meshless method for simulating incompressible fluids on
  \uppercase{GPU}.
\newblock In {\em 2017 19th Symposium on Virtual and Augmented Reality (SVR)},
  pages 297--308, Nov 2017.

\bibitem{wald2004an}
Ingo Wald, Andreas Dietrich, and Philipp Slusallek.
\newblock An interactive out-of-core rendering framework for visualizing
  massively complex models.
\newblock In {\em SIGGRAPH '05}, 2004.

\bibitem{overby2002interactive}
D.~{Overby}, Z.~{Melek}, and J.~{Keyser}.
\newblock Interactive physically-based cloud simulation.
\newblock In {\em 10th Pacific Conference on Computer Graphics and
  Applications, 2002. Proceedings.}, pages 469--470, 2002.

\bibitem{fernandes2013phd}
Davi~T. Fernandes.
\newblock {\em Implementa\c{c}\~{a}o de framework computacional de
  paraleliza\c{c}\~{a}o h\'{i}brida do Moving Particle Semi-implicit Method
  para modelagem de fluidos incompress\'{i}veis}.
\newblock PhD thesis, Universidade de S\~{a}o Paulo, 2013.

\bibitem{fernandes2015domain}
Davi~Teodoro Fernandes, Liang-Yee Cheng, Eric~Henrique Favero, and Kazuo
  Nishimoto.
\newblock A domain decomposition strategy for hybrid parallelization of moving
  particle semi-implicit ({MPS}) method for computer cluster.
\newblock {\em Cluster Computing}, 18(4):1363--1377, Dec 2015.

\bibitem{tompson2016accelerating}
Jonathan Tompson, Kristofer Schlachter, Pablo Sprechmann, and Ken Perlin.
\newblock Accelerating eulerian fluid simulation with convolutional networks.
\newblock {\em arXiv preprint arXiv:1607.03597}, 2016.

\end{thebibliography}







\end{document}